\documentclass[useAMS,usenatbib]{mn2e}
\usepackage{graphicx,amsmath,txfonts}

\title{Protoplanetary migration in non-isothermal discs with turbulence driven by stochastic forcing}
\author[A. Pierens, C. Baruteau and F. Hersant]{A. Pierens $^{1,2},$  C. Baruteau $^3$, F. Hersant  $^{1,2}$\\
$^1$Universit\'e de Bordeaux, Observatoire Aquitain des Sciences de l'Univers,
    BP89 33271 Floirac Cedex, France \\
$^2$   Laboratoire d'Astrophysique de Bordeaux,
     BP89 33271 Floirac Cedex, France \\
$^3$ DAMTP, University of Cambridge, Wilberforce Road, Cambridge CB30WA, United Kingdom}
\date{Released 2012 Xxxxx XX}

\pagerange{\pageref{firstpage}--\pageref{lastpage}} \pubyear{2002}

\def\LaTeX{L\kern-.36em\raise.3ex\hbox{a}\kern-.15em
    T\kern-.1667em\lower.7ex\hbox{E}\kern-.125emX}
\begin{document}
\label{firstpage}
\maketitle
\begin{abstract}
Low-mass objects embedded in isothermal protoplanetary discs are known to suffer rapid inward Type I migration.   In non-isothermal discs, recent work has shown that a  decreasing radial profile of the disc entropy can lead to a strong positive 
corotation torque which can significantly slow down or reverse Type I migration in laminar viscous disc models, depending on the
amount of viscous and thermal diffusion operating  in the planet's horseshoe region. Since the latter is a fraction 
of the pressure scale height of the disc, it is not clear however how this picture changes 
in turbulent disc models.
%{
%Turbulence due to the magneto-rotational instability (MRI) is thought to be the main source of angular momentum transport in protoplanetary discs. Type I migration of protoplanets arising from tidal interaction 
%with their parent disc may be strongly affected by the generated turbulent fluctuations. 
%}
The aim of this study is to examine the impact of turbulence on the torque experienced by a low-mass planet 
embedded in a non-isothermal protoplanetary disc. We particularly focus on the role of turbulence on the 
corotation torque whose amplitude depends on the efficiency of diffusion processes in the planet's horseshoe 
region. The main issues we want to address are whether the part of the corotation torque scaling with the entropy 
gradient can remain
unsaturated in the presence of turbulence and whether the saturation process in non-isothermal discs
can be satisfactorily modelled using laminar disc models.
We performed 2D numerical simulations using a grid-based hydrodynamical code in which turbulence is modelled as stochastic forcing. In order to provide estimations for the viscous and thermal diffusion coefficients 
as a function of the amplitude of turbulence,  we first set up non-isothermal disc models for different 
values of the amplitude of the turbulent forcing. We then include a low-mass planet and determine the 
evolution of its running time-averaged torque.
We show that in non-isothermal discs, the entropy-related corotation torque can indeed remain unsaturated in 
the presence of turbulence.  For turbulence amplitudes that do not strongly affect the disc temperature profile, 
 we find that the running time-averaged torque experienced  by an embedded  
protoplanet is in  fairly good agreement with laminar disc models with appropriate values 
for the thermal and viscous diffusion coefficients and with the formulae of Paardekooper et al. (2011) for the 
total torque in non-isothermal discs. In discs with turbulence driven by stochastic forcing, the corotation torque therefore behaves  similarly as in laminar viscous discs and can 
be responsible for significantly slowing down or reversing Type I migration. 
\end{abstract}

\section{Introduction}
In the course of its evolution, the tidal interaction of a planet with the protoplanetary disc in which 
it formed is known to cause a significant change of the planet orbital elements.  In particular, low-mass planets 
can experience dramatic changes of their distance from the central star under 
the process of Type I migration (Ward 1997; Tanaka et al. 2002). Type II migration concerns more massive planets that are 
able to strongly modify the underlying disc structure by creating a gap around their orbit 
(Lin \& Papaloizou 1986). In that case, 
a planet undergoing Type II migration is locked to the disc viscous evolution and migrates inward on a timescale 
corresponding to the disc viscous timescale, provided that the characteristic disc mass is still larger than the planet mass (Nelson et al. 2000).\\
The gravitational torque exerted by the disc on the planet and causing Type I migration  consists of two components. The differential 
Lindblad torque results from the angular momentum exchange between the planet and the spiral density waves it 
  generates inside the disc.  For sufficiently low-mass planets, the Lindblad torque is well described by 
a linear analysis, which shows that the specific value of this torque scales linearly with disc mass and planet mass and as the inverse square of the disc aspect ratio (Tanaka et al. 2002). Although its sign depends on the  density and temperature gradients inside the  disc, the differential Lindblad 
torque is generally negative for typical disc models and is therefore responsible for inward migration.\\
The corotation torque is due to the torque exerted by the material located in the coorbital region of the planet. A linear 
analysis reveals that the corotation torque consists of a barotropic part  which scales with the vortensity (i.e. 
the ratio between the vertical component of the vorticity and the disc surface density) gradient 
 (Goldreich \& Tremaine 1979) plus an entropy-related part which scales with the entropy gradient (Baruteau \& Masset 2008; Paardekooper \& Papaloizou 2008). 
A negative vortensity (resp. entropy) gradient gives 
rise to a positive vortensity (resp. entropy) related corotation torque. 
%However, Paardekooper \& Papaloizou (2009) have shown 
%that except at early times or in the limit of strong diffusion processes, such a linear approximation breaks down 
%at all protoplanetary masses. In that case, the corotation torque is replaced by the horseshoe drag which results in an 
%isothermal disc from 
%angular momentum exchange between the planet and the disc fluid elements performing horseshoe orbits in the frame 
%rotating with the planet (Ward 1991; Masset \& Casoli 2009; Paardekooper et al. 2010). Similarly to the linear corotation torque, the horseshoe drag scales with both the vortensity 
%and entropy gradients across the horseshoe region but is found to be much more larger. 
The consequence is that for midly 
positive surface density gradients or negative entropy gradients, a positive  corotation torque can eventually counteract the effect of a
negative differential Lindblad torque,  which may stall or even reverse migration (Masset et al. 2006; Paardekooper \& Papaloizou 2009). 
 In isothermal discs, the corotation torque is a non-linear process generally referred as the horsehoe drag and whose amplitude 
is controlled by advection-diffusion of vortensity inside the horsehoe region. In non-isothermal discs, singular production 
of vortensity also arises due to the entropy discontinuity on downstream separatrices (Masset \& Casoli 2009; 
Paardekooper et al. 2010) and in that case, the amplitude of the 
corotation torque is therefore powered by advection-diffusion-production of vortensity inside the horseshoe region.\\
In the absence of any diffusion processes inside the disc, vortensity and entropy gradients 
 across the horseshoe region tend to flatten through phase mixing, which causes the two components of the 
horseshoe drag to  saturate. 
Consequently, desaturating the horseshoe drag requires that some amount of viscous and thermal diffusions are operating inside the 
horseshoe region. In that case, the amplitude of the horseshoe drag depends on the ratio between the diffusion timescales and the horseshoe 
libration timescale and its optimal value, also referred as the fully unsaturated horseshoe drag,  
is obtained when the diffusion timescales are approximately equal to half the horseshoe libration time 
 (see e.g. Baruteau \& Masset 2012 for a recent review). 
In the limit where the diffusion timescales become shorter than the U-turn timescale, the corotation torque decreases and approaches the value 
predicted by linear theory. Therefore, the corotation torque can be considered as a linear combination of the fully unsaturated 
horseshoe drag and the linear corotation torque with coefficients depending on the ratio between the diffusion timescales and 
the horseshoe libration timescale. Corotation torque 
formulae as a function of viscosity and thermal diffusivity were recently proposed by Paardekooper et al. (2011) and 
Masset \& Casoli (2010). \\
So far, most of the studies aimed at studying the process of saturation of the corotation torque focused on laminar disc models with prescribed values of kinematic viscosity and thermal diffusivity. However, it is likely that both angular momentum and heat transport in protoplanetary discs occur because of turbulence probably due to the magneto-rotational instability (MRI, Balbus \& Hawley 1991) and it is not 
clear how the disc turbulence really impacts the corotation torque. In the case of isothermal discs, Uribe et al. (2011) 
performed 3D MHD simulations of planet migration in stratified discs that suggested the torques coming from 
the corotation region can remain unsaturated due to viscous stresses in the disc. Baruteau et al. (2011) recently 
confirmed the existence of an unsaturated corotation torque and demonstrated that a horseshoe dynamics is indeed at work in turbulent discs 
with a weak toroidal magnetic field. Using 2D hydrodynamical simulations in which disc turbulence is modelled through stochastic 
forcing, Baruteau \& Lin (2010) showed that in isothermal discs, the running time-averaged torque experienced by a 
protoplanet is in good agreement with that obtained from a laminar disc model with a similar vortensity diffusion coefficient, and  in which the total torque consists of  the Lindblad torque plus the  barotropic part of the corotation torque. \\
In this paper we adopt a similar strategy as in Baruteau \& Lin (2010) but focus on non-isothermal discs in which the horseshoe 
drag has an additional contribution due to the presence of an entropy gradient across the horseshoe region. The aim 
is to investigate whether the entropy-related horseshoe drag remains unsaturated in presence of turbulence and whether the running
 time-averaged torques in turbulent runs are in good agrement with a laminar viscous disc model with appropriate values for the vortensity and 
entropy diffusion coefficients. \\
% and in which the torque exerted on a planet is the sum of the differential Lindblad torque and 
%of the entropy-related corotation torque. 
In the context of planetary population synthesis models aimed at reproducing 
the  mass-semi-major axis diagram of exoplanets, it is of crucial importance to investigate whether in turbulent disc
 models the entropy-related corotation torque behaves as in laminar viscous discs since it may be responsible for significantly 
slowing down or reversing Type I migration in the latter case. \\
%Discrepancies between observational data and current models based on standard expressions for the torque driving Type I 
%migration (Tanaka et al. 2002) indeed arise unless the corresponding  migration rate is lowering by a fraction 
%$f\sim 0.1-0.01$ (Ida \& Lin 2008; Mordasini et al. 2009).  \\
The paper is organized as follows. In Sect. 2, we give the basic equations and notations we use in this work and describe 
the numerical set-up in Sect. 3. In Sect. 4, we present the properties of turbulence in our simulations. In particular, 
we evaluate the vortensity and entropy diffusion coefficients as well as the turbulent Prandtl number. In Sect. 5, we estimate the effects of 
turbulence on the entropy-related corotation torque and compare with the torque formulae of Paardekooper et al. (2011) and 
with laminar calculations as well. We 
finally discuss our results and draw our conclusions in Sect.6. 

\section{Physical model}
\subsection{Basic equations}
We adopt a 2D adiabatic disc model for which all the physical 
quantities are vertically averaged. We work in a frame rotating 
with angular velocity $\Omega_{p}$,
and adopt cylindrical polar coordinates $(R,\phi)$ with the origin
located at the position of the central star. The continuity, equation of motion and 
energy equation respectively read:
\begin{equation}
\frac{\partial{\Sigma}}{\partial t}+\nabla\cdot(\Sigma \vec{v})=0,
\end{equation}
\begin{equation}
\frac{\partial{\vec{v}}}{\partial t}+(\vec{v}.\nabla)\vec{v}=-\frac{\nabla p}{\Sigma}
-\nabla \Phi,
\label{eq:vel}
\end{equation}
and, 
\begin{equation}
\frac{\partial{S}}{\partial t}+(\vec{v}\cdot\nabla) S=0
\end{equation}
where $\Sigma$ is the disc surface density, $\vec{v}$ the velocity, $P$ the pressure and 
$S$ the entropy which is defined as:
\begin{equation}
S=\frac{p}{\Sigma^{\gamma_{ad}}},
\end{equation}
where $\gamma_{ad}$ is the adiabatic index. Here we use an ideal gas equation of state 
$p=\Sigma R_g T/\mu $ where $R_g$ is the gas constant, T the temperature and $\mu$ 
the mean molecular weight. \\
In Eq. \ref{eq:vel}, $\Phi$ is the gravitational potential and reads:
\begin{equation}
\Phi=-\frac{GM_\star}{R}-\frac{1}{2}\Omega_{p}^2 R^2+\Phi_{ind}+\Phi_{turb}
\end{equation}
where $M_\star$ is the mass of the central star and  $\Phi_{ind}$ is an indirect term which accounts from the fact that the frame centered 
on the star is non-inertial (e.g. Nelson et al. 2000). Effects resulting from turbulence  
are modelled by applying the external turbulent potential $\Phi_{turb}$ of Laughlin et al. (2004) (see Sect. 3). Previous 
studies (e.g. Baruteau \& Lin 2010; Pierens et al. 2011) have shown that the perturbations induced by this 
stochastic forcing term  have similar  statistical properties to those resulting from 3D MHD simulations. In 
isothermal discs, it is well known that fluctuations arising from turbulence generate transport of angular momentum through Reynolds and Maxwell stresses (e.g. Papaloizou \& Nelson 2003). For non-isothermal discs, additional turbulent energy transport is expected as a consequence of 
entropy perturbations.

\subsection{Averaged quantities and turbulent  diffusion coefficients}

To estimate the rate of angular momentum and energy transport resulting from turbulence, we adopt a common procedure and
  make use of averaged quantities. The mass-weighted azimuthal average of the quantity $Q(r,\phi,t)$ is given by:
\begin{equation}
\bar Q(R,t)=\frac{\int \Sigma(R,\phi,t) Q(R,\phi,t) d\phi}{\int \Sigma(R,\phi,t) d\phi}
\end{equation}
where the integral is performed over $2\pi$ in azimuth. Setting $v_R=\bar v_R +\delta v_R$ and 
$v_\phi=\bar v_\phi +\delta v_\phi$, where $\delta v_R$ and $\delta v_\phi$ are the radial and azimuthal velocity 
fluctuations respectively, and averaging the azimuthal component of the equation 
of motion (Eq. \ref{eq:vel}) gives the following equation for the conservation of the specific angular momentum 
$j=Rv_\phi$ (e.g. Fromang \& Nelson 2006):
\begin{equation}
\frac{\partial}{\partial t} (\Sigma \bar j)+\frac{1}{R} \frac{\partial}{\partial R} (R \Sigma \bar v_R \bar j)=
-\frac{1}{R}\frac{\partial}{\partial R} (R^2 \Sigma \overline{\delta v_R \delta v_\phi})
\end{equation}
The previous equation can be re-written as (Balbus \& Papaloizou 1999):
\begin{equation}
\frac{\partial}{\partial t} (\Sigma \bar j)+\frac{1}{R} \frac{\partial}{\partial R} (R \Sigma \bar v_R \bar j)=
-\frac{1}{R}\frac{\partial}{\partial R} (\Sigma R^2\alpha_R \overline{P}/\overline{\Sigma})
\end{equation}
where  $\alpha_R$ is the  Reynolds stress parameter  which is defined as:
\begin{equation}
\alpha_R=\frac{\bar \Sigma \overline{\delta v_R \delta v_\phi}}{\bar P}
\label{eq:alpha}
\end{equation}

Adopting a similar averaging procedure for the energy equation leads to the following relation for the conservation of entropy:

\begin{equation}
\frac{\partial}{\partial t} (\Sigma \bar S)+\frac{1}{R}\frac{\partial}{\partial R}(R\Sigma \bar v_R \bar S)=
-\frac{1}{R}\frac{\partial}{\partial R}(R\Sigma \overline{\delta v_R\delta S})
\end{equation}
which can be recast as:
\begin{equation}
\frac{\partial}{\partial t} (\Sigma \bar S)+\frac{1}{R}\frac{\partial}{\partial R}(R\Sigma \bar v_R \bar S)=
\frac{1}{R}\frac{\partial}{\partial R}(\kappa R\Sigma \frac{\partial \bar S}{\partial R})
\end{equation}
where $\kappa$ is  the diffusion coefficient associated with the entropy and is given by:
\begin{equation}
\kappa=-\frac{\overline{\delta v_R\delta S}}{\partial \bar S/\partial R}
\label{eq:kappa}
\end{equation}
%Combining Eqs. \ref{eq:nu} and \ref{eq:kappa}, we get the following expression for the turbulent 
%Prandtl number $Pr=\nu/\kappa$:
%\begin{equation}
%Pr=-\frac{2\partial \bar S/\partial R}{3 \Omega}\frac{\overline{\delta v_R \delta v_\phi}}{\overline{\delta v_R\delta S} }
%\end{equation}

 Here, we  note that the previous equation was derived assuming an adiabatic disc model. This neglects 
turbulent heating, which is modelled in our simulations by artificial viscous heating at shocks arising from 
 the turbulence (see Sect. \ref{sec:num}).  Turbulent heating remains small, however, for the turbulence amplitudes 
considered in this study. To estimate the effective thermal diffusion  coefficient in 
the simulations, we therefore simply neglect the turbulent heating term and make use of Eq. \ref{eq:kappa}.

\section{Numerical set-up}
\label{sec:num}
 Simulations were performed with the FARGO and GENESIS numerical codes which solve the 
 equations for the disc on a polar grid. 
These codes employ an advection scheme based on the monotonic transport algorithm 
(Van Leer 1977) and include the FARGO algorithm (Masset 2000) to avoid time step limitation 
due to the Keplerian orbital velocity at the inner edge of the grid. The energy equation 
that is implemented in both codes reads:
\begin{equation}
\frac{\partial e}{\partial t}+\nabla \cdot (e{\bf v})=-(\gamma_{ad}-1)e{\nabla \cdot {\bf v}}+Q^+_{bulk}
\end{equation}
where $e$ is the thermal energy density and $Q^+_{bulk}$ is the heating source term by artificial 
viscosity and provided by shocks arising from turbulent fluctuations. This term is handled by
 adopting a standard von Neumann-Richtmyer artificial bulk viscosity (see Stone \& Norman 1992) with 
$C_2=1.4$, where $C_2$ represents the number of zones over which artificial viscosity will
spread a shock.  Here, no cooling term is included as an additional source term, which means that reaching a thermal equilibrium 
for the disc is not possible and that the temperature profile may be progressively altered by turbulent heating, especially in the case of strong turbulent fluctuations. However, from the results of simulations presented in Sect. \ref{sec:withplanet}, we anticipate 
that using a cooling term is not necessary since for most of the turbulence amplitudes considered in this 
study, turbulence heating has little impact on the temperature profile over the runtime covered by the 
calculations. In fact,  
not considering such a term enables to investigate effects resulting from turbulent diffusion only without continuously 
changing the temperature profile close to the planet. We further note that the results presented in Sect. \ref{sec:withplanet} may depend strongly on the expression employed for the cooling term. \\ 
 
The computational units  that we adopt are such that the unit of mass is the 
central mass $M_*$, the unit of distance is the initial orbital radius
$R_p$  of 
the  planet and the unit of time is $\Omega_p^{-1}=(GM_*/R_p^3)^{-1/2}$.
In the simulations presented here, we use $N_R=512$ radial grid cells
uniformly distributed between $R_{in}=0.4$ and $R_{out}=1.8$, and 
$N_{\phi}=1200$ azimuthal grid cells. Wave-killing zones are employed
 for $R<0.5$ and $R>1.58$ to avoid wave reflections at the disc edges (de Val-Borro et al. 2006).\\

The initial disc surface density profile is $\Sigma=\Sigma_p\times(R/R_p)^{-\sigma}$ with 
$\Sigma_p=5\times 10^{-4}$ and $\sigma=1.5$ by default. For such a disc surface density profile,
the initial vortensity-related part of the corotation torque cancels out so that the corotation torque consists only of its
entropy-related part. The initial disc aspect ratio is $h=h_p\times(R/R_p)^f$ with $h_p=0.03$ and where 
$f$ is the disc flaring index which is set to $f=0.3$ in this work. This corresponds to an initial temperature 
profile decreasing as $R^{-\beta}$ with $\beta=1-2f=0.4$. For an adiabatic index $\gamma=5/3$ which 
corresponds to the value adopted here, the 
initial entropy profile therefore varies as $R^{-\xi}$ with $\xi=\beta-(\gamma-1)\sigma=-0.59$.  As will be justified 
in Sect. \ref{sec:withplanet}, such a positive gradient for the initial 
entropy profile is chosen to minimize the maximum convergence time of the turbulent runs, since both the entropy-related corotation torque and the differential Lindblad torque are negative in that case.

In the simulations presented here, turbulence is modelled by applying at each time-step a turbulent 
potential $\Phi_{turb}$ to the disc (Laughlin et al. 
2004, Baruteau \& Lin 2010) and corresponding to the superposition of $50$ 
wave-like modes. This reads as:

\begin{equation}
\Phi_{turb}(R,\phi,t)=\gamma R^2 \Omega^2\sum_{k=1}^{50}\Lambda_k(R,\phi,t),
\label{eq:phi}
\end{equation} 

with

\begin{equation}
\Lambda_k=\xi_k e^{-\frac{(R-R_k)^2}{\sigma_k^2}}
\cos(m_k\phi-\phi_k-\Omega_k\tilde{t_k})
\sin(\pi \tilde{t_k}/\Delta t_k).
\label{eq:lambda}
\end{equation}

In Eq. \ref{eq:lambda}, $\xi_k$ is a dimensionless constant  parameter randomly 
  sampled with a Gaussian distribution of unit width. 
$R_k$ and $\phi_k$ are, respectively, the radial and
azimuthal initial coordinates of the mode with wavenumber $m_k$,
 $\sigma_k=\pi R_k /4m_k$ is the radial extent of that mode, and
$\Omega_k$ denotes the Keplerian angular velocity at $R=R_k$.  
 Both $R_k$ and $\phi_k$ are randomly  sampled with a uniform 
distribution, whereas $m_k$ is randomly sampled  with a logarithmic
distribution between $m_k=1$ and $m_k=96$. 
Each mode of wavenumber $m_k$ starts at time $t=t_{0,k}$ and 
 terminates when $\tilde{t_k}=t-t_{0,k} > \Delta t_k $,
where $\Delta t_k=0.2\pi R_k /m_k c_s$ denotes the lifetime of mode with wavenumber $m_k$, $c_s$ being the sound speed.  Such a value for 
$\Delta t_k$ yields  an autocorrelation timescale $\tau_c\sim 0.5 T_{orb}$, 
where $T_{orb}$ is the orbital period at $R=1$ (Baruteau \& Lin 2010, and see Sect. \ref{sec:kappa}).\\ 
Following Ogihara et al. (2007), 
 we set $\Lambda_k=0$ if $m_k > 6$  to save computing time.
 As noticed by Baruteau \& Lin (2010), such an assumption is supported by the fact that 
a turbulent mode with wavenumber $m$ has an amplitude decreasing as $\exp(-m^2)$ and a lifetime 
$\propto 1/m$, so that the contribution to the turbulent potential of a high wavenumber turbulent mode  
 is relatively weak.
  \\ 
In Eq. \ref{eq:phi}, $\gamma$ denotes the value of the turbulent
forcing parameter, which controls the amplitude of the stochastic density perturbations.
In the simulations presented here, we used values for $\gamma$ ranging from $10^{-5}$ to 
$2\times 10^{-4}$.  Given that $\gamma$ is related to the dimensionless $\alpha$  
viscosity of Shakura \& Sunyaev (1973)  by the relation 
$\alpha\sim 120(\gamma/h_p)^2$ (see Sect. \ref{sec:kappa}), the latter values for $\gamma$ 
correspond to $\alpha$ ranging from $\sim 10^{-5}$ to $\sim 5 \times 10^{-3}$.  \\
In the following, we first estimate how both the viscous and thermal 
diffusion coefficients induced by turbulence depend on the amplitude of the turbulent 
potential $\gamma$. We then present the results of calculations of a low-mass planet embedded
 in a non-isothermal disc subject to stochastic forcing. We consider different values for 
$\gamma$ and study how the running time-averaged torque deduced from our simulations compare with 
laminar runs and with the analytical formulae of Paardekooper et al. (2011) for the torque due to Lindblad resonances 
and horseshoe drag in the presence of viscous and thermal diffusion.     

\section{Turbulence properties}

\subsection{Estimation of the turbulent thermal diffusivity}
\label{sec:kappa}

\begin{figure}
\centering
\includegraphics[width=0.95\columnwidth]{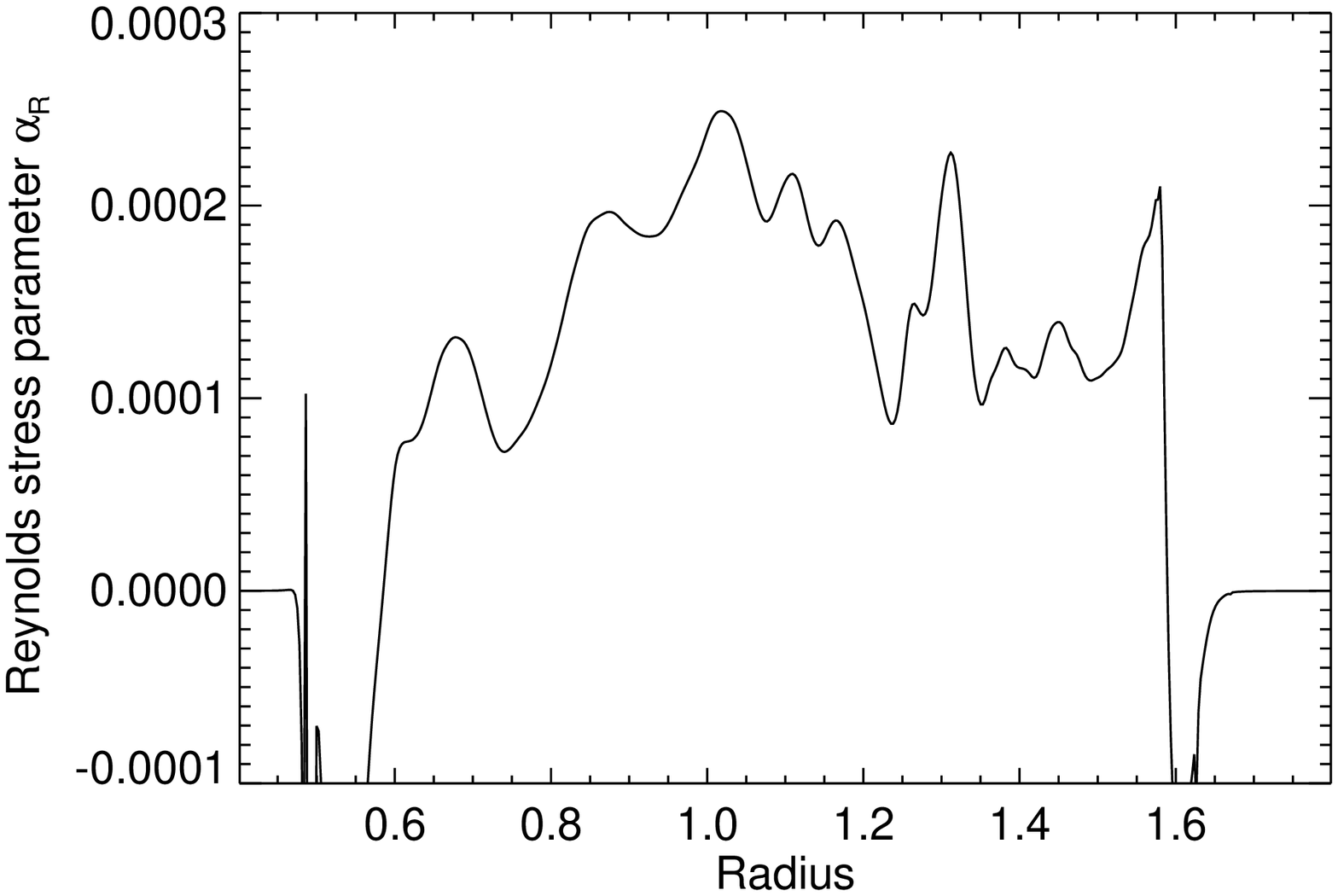}
\includegraphics[width=0.95\columnwidth]{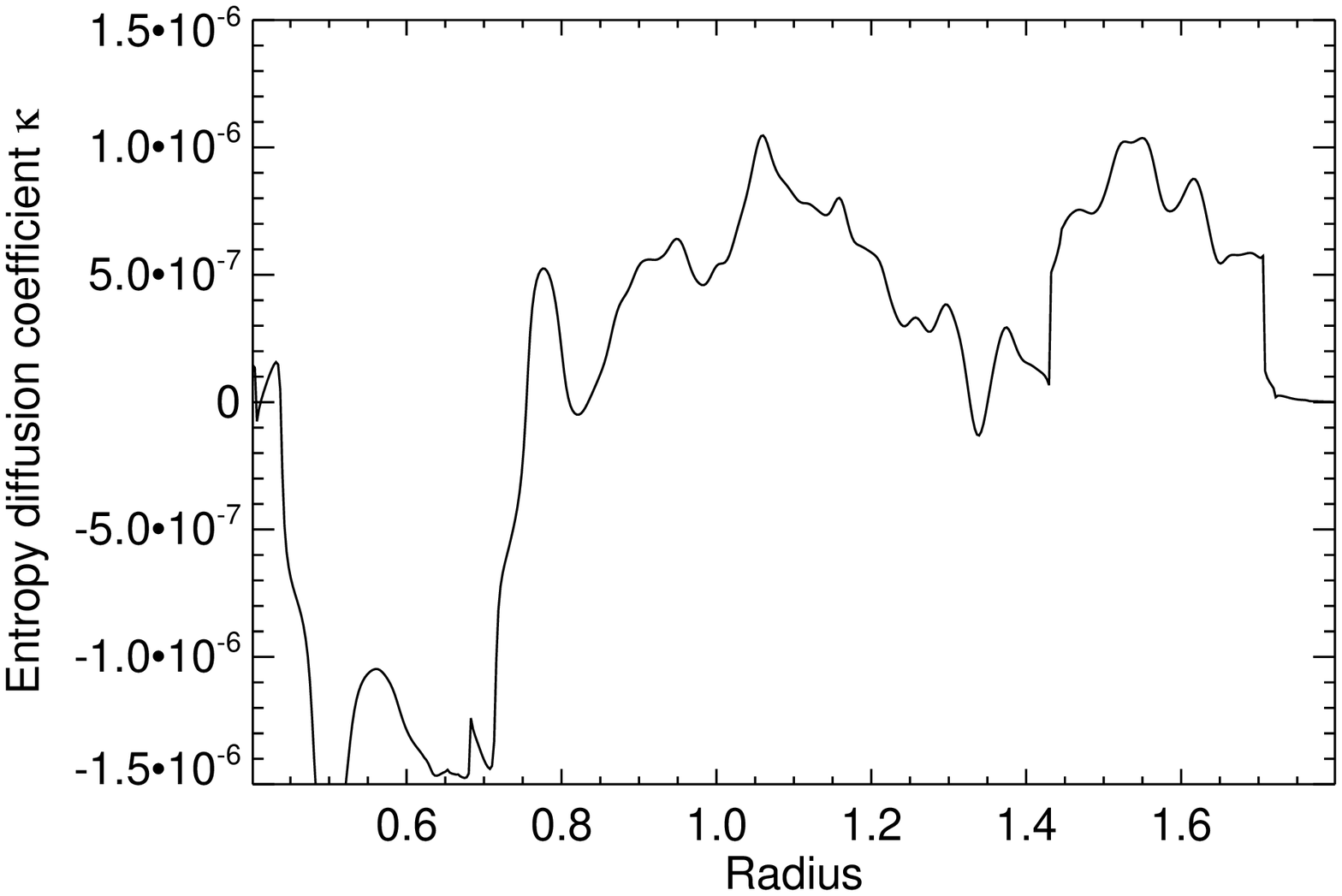}
\caption{{\it Upper panel:} Reynolds stress parameter $\alpha(R)$ as a function of radius for $\gamma=8\times10^{-5}$ 
and time-averaged over $700$ $T_{orb}$. {\it Lower panel:} same but for the entropy diffusion coefficient $\kappa(R)$ defined by Eq. \ref{eq:kappa}.}
\label{fig:kvsr}
\end{figure}
In non-isothermal discs, the main effect of the applied turbulent potential is to generate velocity and temperature 
fluctuations responsible for  both entropy and angular momentum transport. In a previous 
study dedicated to isothermal discs, Baruteau \& Lin (2010) showed that for the adopted turbulent potential, 
the resulting Reynolds stress parameter scales as $\gamma^2$.
Here, in order to investigate the dependence of the entropy diffusion coefficient on the amplitude of 
turbulence $\gamma$, we performed a series of simulations in the non-rotating frame of a non-isothermal disc with turbulence
driven by stochastic forcing, 
varying the value for $\gamma$ from $\gamma=2\times 10^{-5}$ to $\gamma=2\times 10^{-4}$ and using the disc 
model described in Sect. \ref{sec:num}. From the results of these simulations, the Reynolds stress parameter 
$\alpha_R(R,t)$ and the entropy diffusion coefficient $\kappa (R,t)$  are computed using Eqs. \ref{eq:alpha} and 
\ref{eq:kappa} respectively.  To examine how these two quantities vary with radius,  a time 
average over $700$ $T_{orb}$ is performed. For $\gamma=8\times 10^{-5}$, Fig. \ref{fig:kvsr} displays the 
time-averaged coefficients $\alpha_R(R)$ 
and $\kappa(R)$ as a function of radius. Clearly, both quantities appear to be
approximately uniform with radius, except in the vicinity of the wave-killing zones.  \\

\begin{figure}
\centering
\includegraphics[width=0.95\columnwidth]{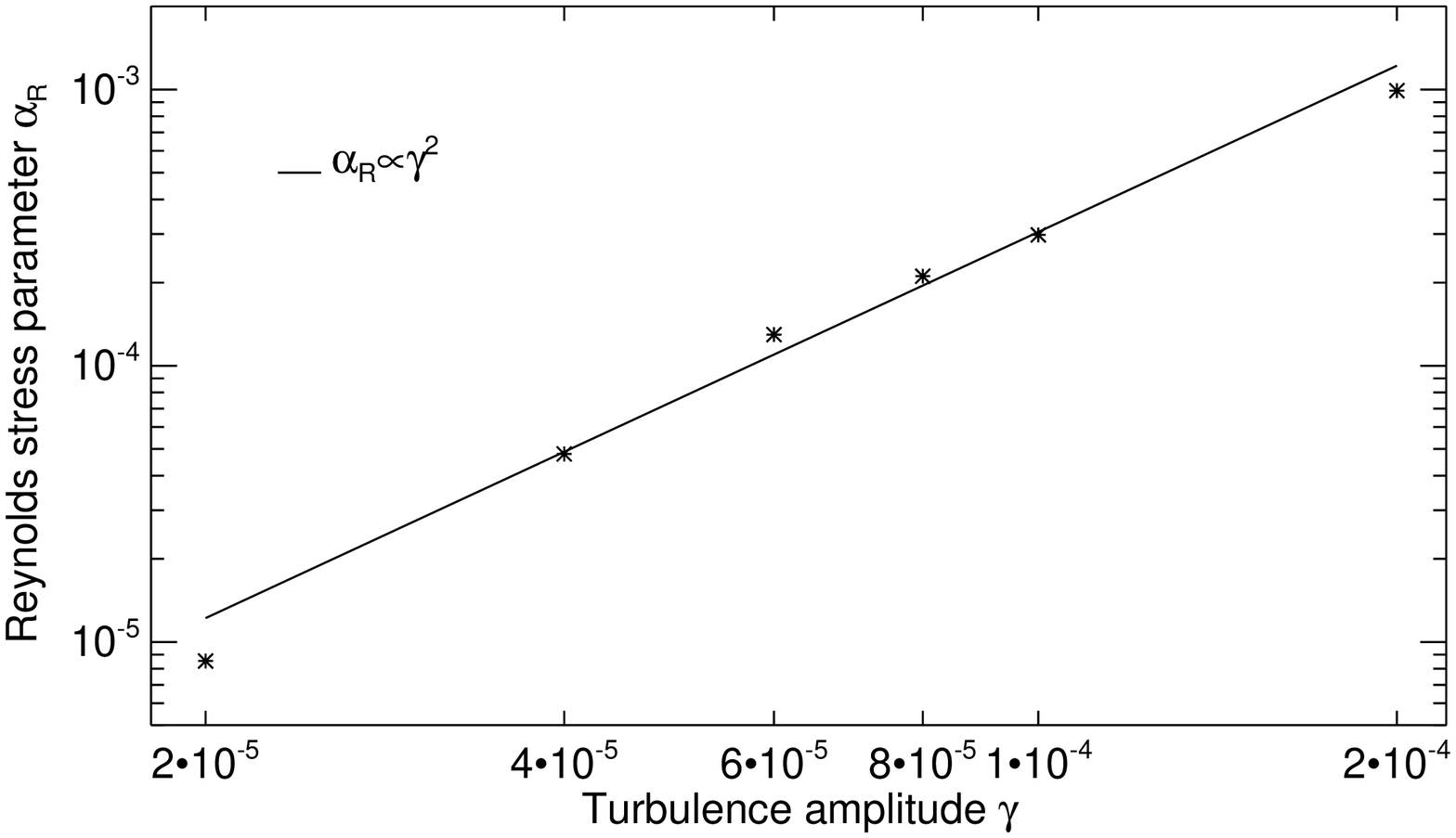}
\includegraphics[width=0.95\columnwidth]{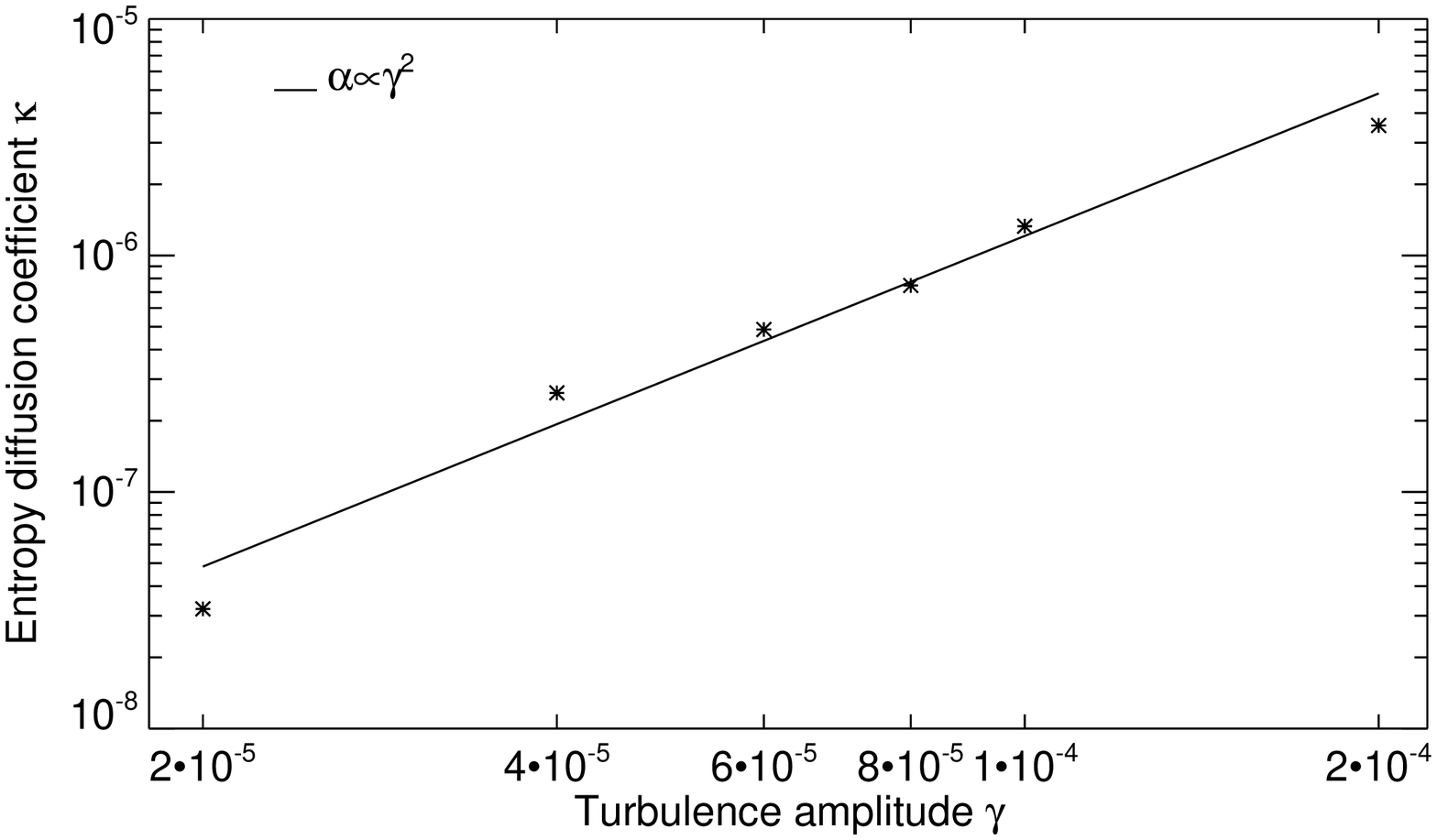}
\caption{{\it Upper panel:} Averaged Reynolds stress parameter $\alpha$ as a function of turbulence amplitude $\gamma$.
{\it Lower panel:} Averaged entropy diffusion coefficient $\kappa$ as a function  of turbulence amplitude $\gamma$.}
\label{fig:akvsg}
\end{figure}

We now consider an additional space average of $\alpha_R(R)$ and $\kappa(R)$ in the range $0.9\le R \le 1.1$ and 
 in the following, we simply denote 
by $\alpha_R$ and $\kappa$  the resulting numbers.  In  simulations with an embedded planet presented in Sect. 
\ref{sec:withplanet}, the planet is on a fixed circular orbit located at $R_p=1$. Given that the corotation torque 
experienced by the planet depends on the viscous and thermal diffusion timescales across the horseshoe region whose 
half-width is a fraction of the disc scale height $H$ for the planet masses we consider,  the range over which the space average 
is performed therefore corresponds to the disc region which contributes the most to the torque exerted on the planet. The upper  panel of Fig. \ref{fig:akvsg} represents 
$\alpha_R$ as a function of $\gamma$. As expected,  $\alpha_R$ appears to scale as $\gamma^2$ 
and is found to be well approximated by:
\begin{equation}
\alpha_R \; \approx 30 \left(\frac{\gamma}{h_p}\right)^2, 
\label{eq:fitalpha}
\end{equation}
which is in  excellent agreement with the relation derived by Baruteau \& Lin (2010). 
A slight discrepancy arises because we average with radius in the range $0.9 \le R \le 1.1$ whereas 
the space average is performed over the  disc region located outside the wave-killing zones  in Baruteau \& Lin (2010). 
 While in laminar discs vortensity diffusion is controlled by the kinematic viscosity, this is not necessary the 
case in turbulent discs
where the vortensity's diffusion coefficient $D$  can differ from the turbulent viscosity 
$\nu_R=\alpha_R c_s H$  associated with the turbulent diffusion of angular momentum. For 
instance, Baruteau \& Lin (2010) 
have shown that in isothermal discs, the ratio $S_c=\nu_R/D$ usually referred as the Schmidt number is
$S_c\sim 0.25$. Although not shown here, we have checked that this value also holds in non-isothermal discs and in the following, 
we will make use of the dimensionless vortensity's diffusion coefficient that we define as 
$\alpha_D=S_c^{-1}\alpha_R\sim 4 \alpha_R$. Given that in MHD calculations, the  total viscous stress parameter 
corresponding to the dimensionless $\alpha$ viscosity of Shakura \& Sunyaev (1973) is 
$\sim 4 \alpha_R$ (Fromang \& Nelson 2009),  this suggests that our value for $\alpha_D$ is close to that 
corresponding to the standard $\alpha$ viscous stress parameter. Using Eq. \ref{eq:fitalpha}, the dimensionless 
viscosity of Shakura \& Sunyaev (1973) can consequently be well approximated  by:
\begin{equation}
\alpha \; \approx 120 \left(\frac{\gamma}{h_p}\right)^2, 
\label{eq:fitalphav}
\end{equation}

We plot the entropy diffusion coefficient $\kappa$ as a function of $\gamma$ in the lower panel of Fig. 
\ref{fig:akvsg}. 
Again, $\kappa$ is found to increase as $\gamma^2$, which is in agreement with the expectation that both $\delta v_r$ and 
$\delta S$ scale with $\gamma$. The best-fitting solution is found to be:
\begin{equation}
\kappa \; \approx 125 \; \gamma^2 R_p^2\Omega_p,
\label{eq:fitkappa} 
\end{equation}
\begin{figure}
\centering
\includegraphics[width=0.49\columnwidth]{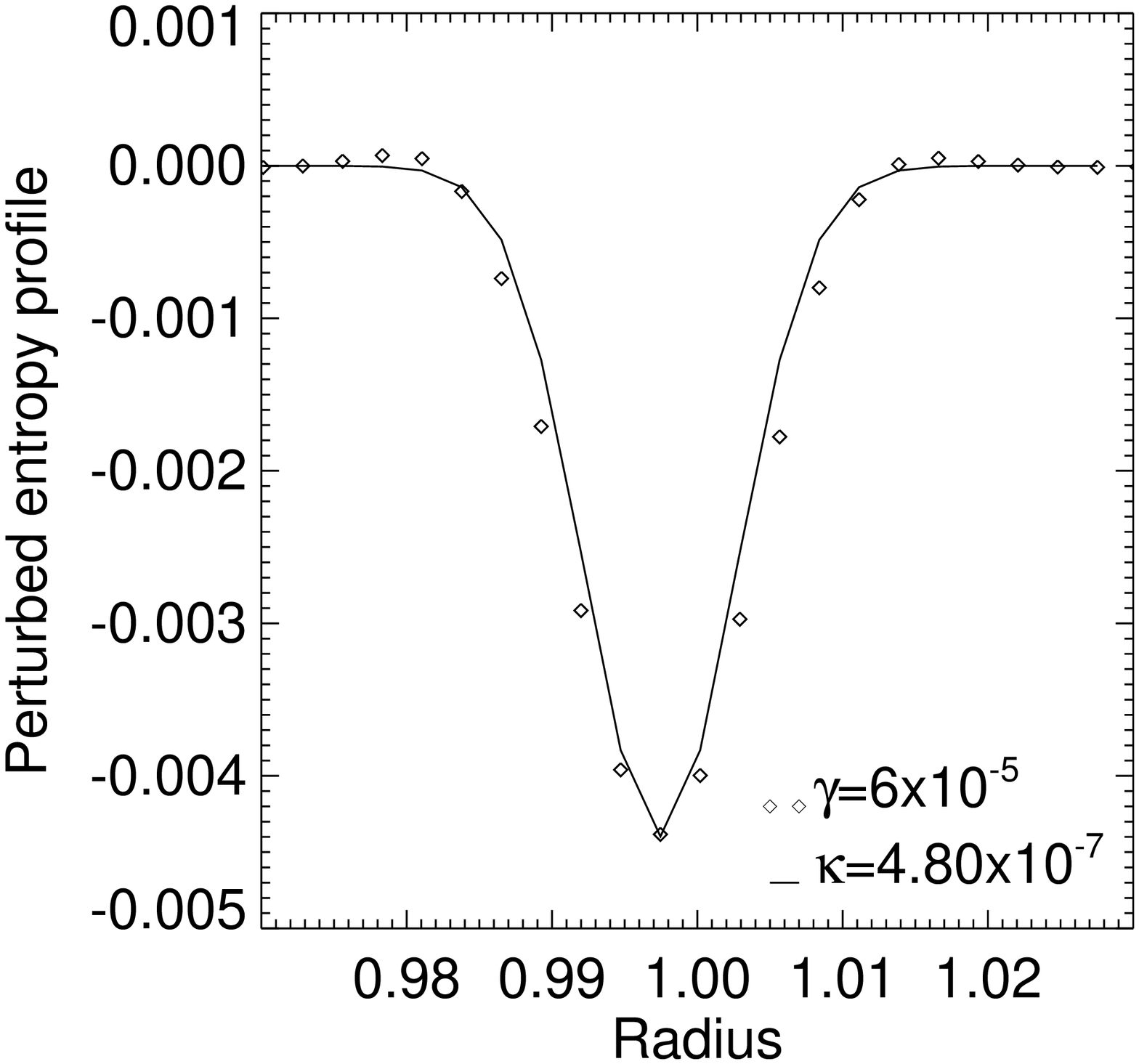}
\includegraphics[width=0.49\columnwidth]{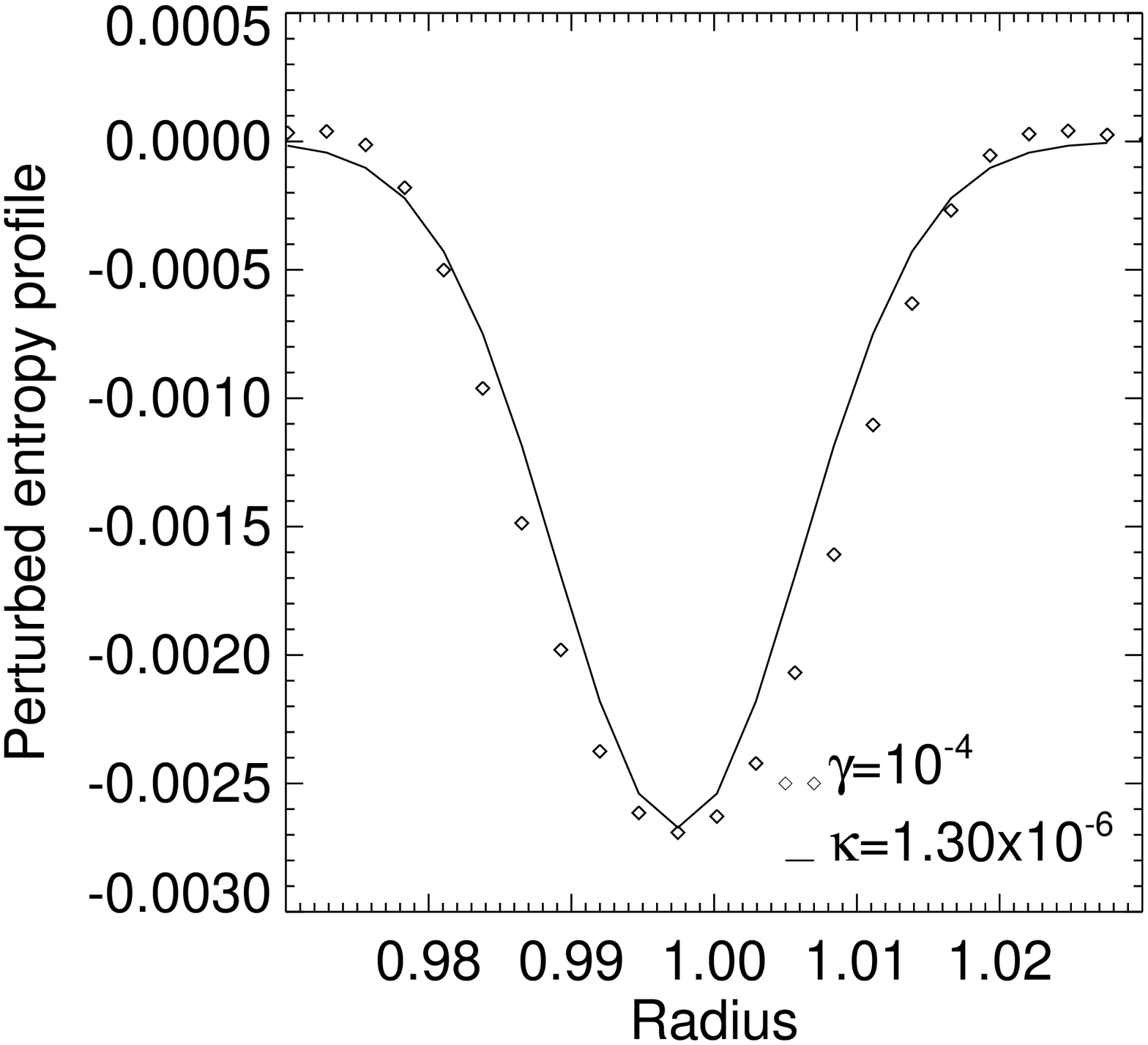}
\caption{Perturbed entropy profile at $t=4.5T_{orb}$  averaged over six different realizations  with $\gamma=6\times 10^{-5}$ 
({\it left panel}) and 
$\gamma=10^{-4}$ ({\it right panel}) and in which a $10 \%$ negative perturbation is applied to the 
background entropy profile at $R_d=1$. The solid line corresponds to the analytical fit given by 
Eq. \ref{eq:gaussfit} with $\kappa$  given by Eq. \ref{eq:fitkappa}.}
\label{fig:gaussian}
\end{figure}

\begin{figure}
\centering
\includegraphics[width=0.95\columnwidth]{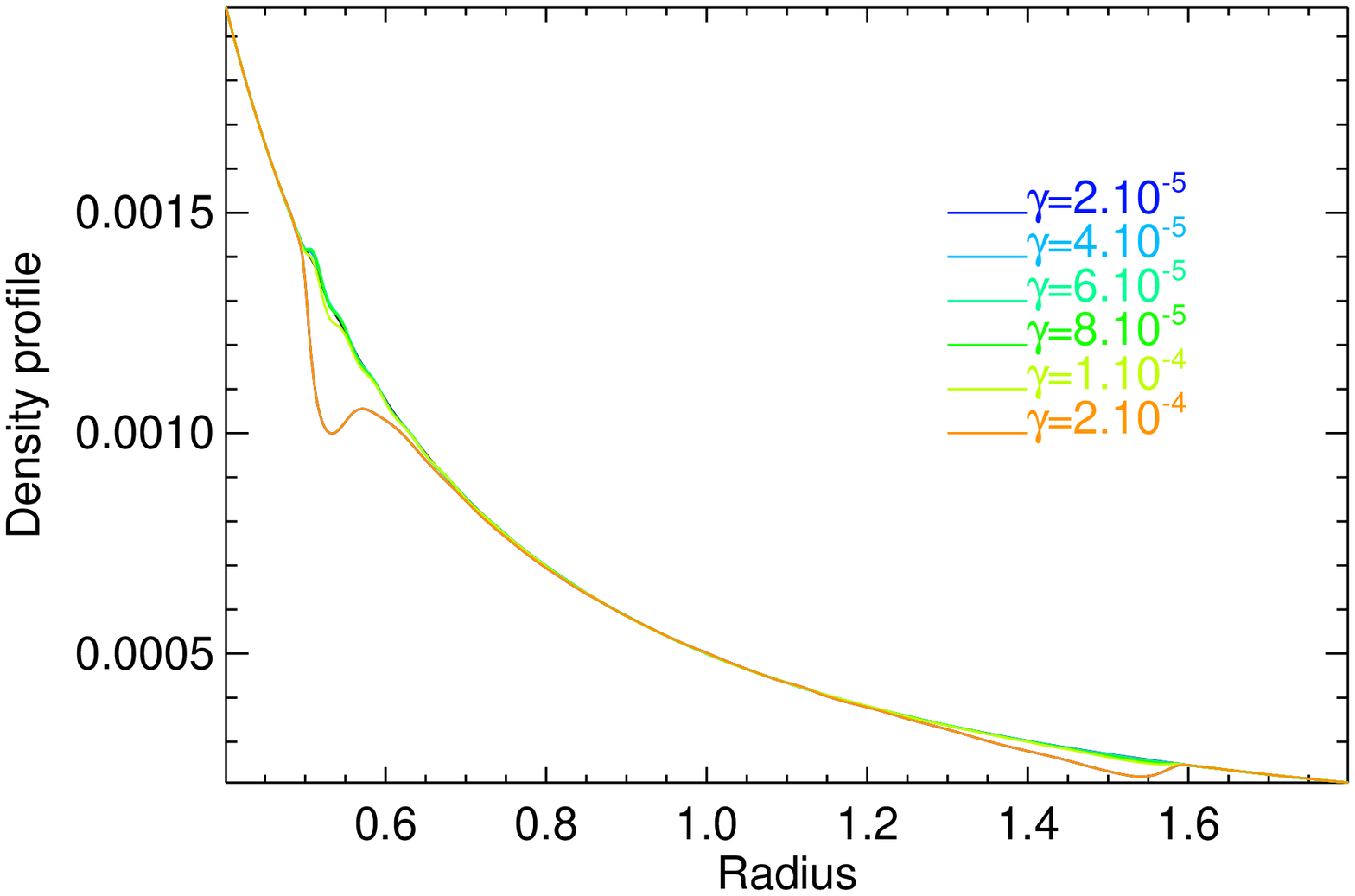}
\includegraphics[width=0.95\columnwidth]{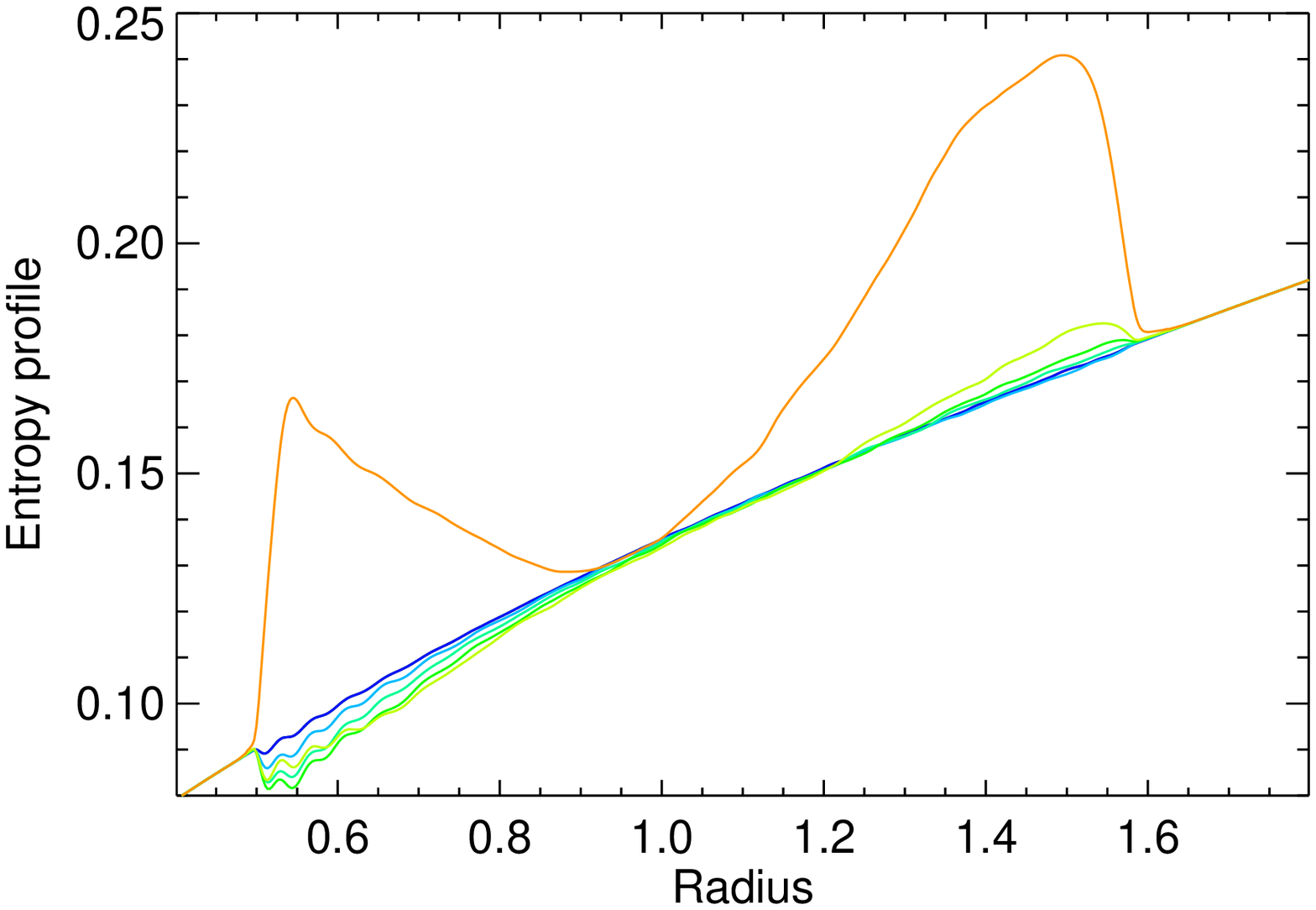}
\caption{{\it Upper panel:} Surface density profile at $t=700 \; T_{orb}$ for the different values of $\gamma$ we 
considered. {\it Lower panel:} Same but for the entropy profile.}
\label{fig:profiles}
\end{figure}

To check the consistency of the previous estimation, we followed the approach of Baruteau \& Lin (2010) and 
performed a series of runs varying the value for $\gamma$ 
and in which a $10 \%$ negative perturbation with a $\delta$-function profile is applied to the initial entropy profile at 
$R_d\sim 1$. This is achieved by applying a positive perturbation to the density profile together with 
a negative temperature perturbation with same amplitude, in such a way that the initial pressure perturbation is zero. 
The evolution of the perturbed entropy profile is then compared with  the function:
\begin{equation}
\delta S(R,t)=\frac{\delta S_0\Delta R}{\sqrt{4\pi \kappa t}} \exp\left(\frac{-(R-R_d)^2}{4\kappa t}\right),
\label{eq:gaussfit}
\end{equation}
where $\delta S_0$ is the initial entropy perturbation at $R=R_d$, $\Delta R$ is the radial resolution of the grid 
and $\kappa$ is the estimation of the entropy diffusion coefficient given by  Eq. \ref{eq:fitkappa}. 
The results of this set of simulations are presented in Fig. \ref{fig:gaussian} where we plot 
for $\gamma=6\times 10^{-5}$ and $\gamma=10^{-4}$ 
both the perturbed entropy profile averaged over six different realizations at $t=4.5$ $T_{orb}$ and the fitting formula given 
by Eq. \ref{eq:gaussfit}. Clearly, the 
perturbed entropy profile closely matches the expected one, which indicates that 
Eq. \ref{eq:fitkappa} provides a fairly good estimation of the effective entropy diffusion coefficient.\\ 
It is worthwhile to notice that this model can suffer from the time evolution of the entropy profile due to artificial viscous heating at shocks. This is illustrated in Fig. \ref{fig:profiles} which shows the surface density and entropy profiles 
at $t=700\; T_{orb}$ for the different values of $\gamma$ we considered. Although the surface density profile 
remains almost constant, we see 
that the entropy profile can be significantly altered in turbulent runs with high values of $\gamma$. 
 This is due not only to turbulent diffusion which makes the 
entropy profile steeper but also to  turbulent heating  which tends 
to increase the local value for the entropy. Looking at Fig. \ref{fig:profiles}, we see that effects resulting from turbulent 
heating are more pronounced near the radial boundaries. 
%This arises because the initial entropy profile 
%is restored in the wave-killing zones while the entropy profile progressively deviates from the initial 
%one in the active part of the disc. Consequently, a strong entropy gradient can be formed at the disc 
%edges which enhances turbulent heating process by artificial viscosity.} . 
However, we are 
confident that this process does not strongly affect our estimations of $\alpha$ and $\kappa$ since these are determined
using space averages in the range $0.9<R<1.1$, which corresponds to disc regions where the entropy profile is only 
slightly modified over the runtime of the simulations. To check this, we computed these two quantities 
using normalizations by the initial pressure or entropy profiles 
and found good agreement with the estimations given by Eqs. \ref{eq:fitalphav} and \ref{eq:fitkappa} for which 
normalizations by time averaged profiles were considered.

\subsection{Determination of the turbulent Prandtl number}

In this section, we investigate how the turbulent Prandtl number $P_r$ depends on the amplitude of the turbulent  
forcing $\gamma$. The turbulent Prandtl number is commonly defined as:

\begin{equation} 
P_r=\frac{\nu_R}{\kappa}=\frac{\alpha_Rc_s H}{\kappa}
\label{eq:prandtl}
\end{equation}

\begin{figure}
\centering
\includegraphics[width=0.95\columnwidth]{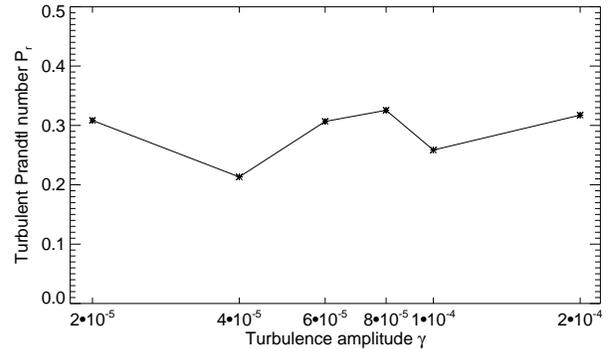}
\caption{Estimated turbulent Prandtl number as a function of turbulence amplitude $\gamma$.}
\label{fig:prandtl}
\end{figure}

%{\bf We note that as defined here, $Pr$ do not correspond to the ratio between the vortensity and entropy diffusion 
%coefficients since it was shown in Baruteau \& Lin (2010) that $\nu$ differs from the vortensity diffusion by a factor $Sc$ where $Sc\sim 0.25$ is the 
%Schmidt number.}
Since both $\alpha_R$ and $\kappa$ scale with $\gamma^2$, we expect $P_r$ to not depend on the value 
for $\gamma$. This is confirmed by inspecting Fig. \ref{fig:prandtl} where $P_r$ is displayed  as 
as a function of $\gamma$ and which reveals that the turbulent Prandtl number is $P_r\sim 0.3$ on average. 
Such a  value 
is close  to that derived by Ruediger et al. (1988) who estimate that a value $P_r\sim 0.1$ is necessary to explain the outburst behaviour of  cataclysmic variables. In the case of shear-driven hydrodynamic turbulence, it is also of interest to mention that Lathrop et al. (1992) measured a turbulent Prandtl number $P_r\sim 0.176$ 
in experiments of Couette-Taylor flows at Reynolds numbers well beyond the onset of chaos.

\subsection{Turbulent transport of entropy in a radiative protoplanetary disc}

As mentioned above, no cooling source term is included in the energy equation that is 
employed in this work. However, in a more realistic disc model in which turbulent effects and 
radiative processes are both included, we expect a thermal equilibrium state to be reached  
when radiative cooling compensates turbulent heating. For such a radiative turbulent disc,  diffusion of entropy is due   
not only to turbulent transport  as it is the case for the idealized disc considered here  but also to radiative diffusion. For an active laminar disc 
model with turbulent heating balanced by radiative losses, the  thermal diffusion 
coefficient $\kappa_{rad}$ associated with radiative transport can be expressed as (Paardekooper et al. 2011):
\begin{equation}
\kappa_{rad}=9\gamma_{ad}(\gamma_{ad}-1)\nu\left(1+\frac{2\sqrt{3}}{3\tau}+\frac{2}{3\tau^2}\right) 
\label{eq:kappar}
\end{equation}
where $\tau$ is the optical depth and $\nu=\alpha c_s H$ the kinematic viscosity. Noting that 
$\nu=S_c^{-1}P_r \; \kappa$ from the results of the previous section and using Eq. \ref{eq:kappar}, 
we obtain the following expression for the ratio between the thermal diffusion coefficient 
associated with radiative effects and that associated with the turbulent transport of entropy:
\begin{equation}
\frac{\kappa_{rad}}{\kappa}=9S_c^{-1}P_r\gamma_{ad}(\gamma_{ad}-1)\left(1+\frac{2\sqrt{3}}{3\tau}+\frac{2}{3\tau^2}\right) 
\label{eq:ratio}
\end{equation}
Previous equation shows than in optically thick inner regions of protoplanetary discs, turbulent and radiative diffusion 
coefficients will tend to be of the same order of amplitude whereas in optically thin outer regions, radiative transport 
will be much more efficient than turbulent transport for diffusing entropy.

\section{Effects of turbulence on the entropy-related corotation torque}
\label{sec:withplanet}

In this section, we evaluate the  effect of turbulence on the entropy-related corotation torque. 
To investigate this issue, we performed simulations of a low-mass planet on a 
circular orbit at $R_p=1$ and $\phi_p=\pi$ and embedded in a non-isothermal 
disc subject to stochastic forcing. We used values for the amplitude of the turbulent potential 
ranging from $\gamma=10^{-5}$ to 
$\gamma=2\times 10^{-4}$. The disc model is the same as described in Sect. \ref{sec:num} 
with  power-law indexes for 
the initial surface density, temperature and entropy profiles of $\sigma=1.5$, $\beta=0.4$ and $\xi=-0.59$  
respectively and an aspect ratio at the planet position of $h_p=0.03$.
We considered planets with mass ratio 
$q=5\times 10^{-6}$ and $q=10^{-5}$ which corresponds to planets with masses of $1.6$ $M_\oplus$ and 
$3.2$ $M_\oplus$ if the central star has a solar mass.  For $q=5\times 10^{-6}$ and for a 
softening parameter $b=0.4h_p$ which is the value set in this work, the half-width of the 
horseshoe region is  $x_s\sim 1.1/{\gamma_{ad}}^{1/4} R_p \sqrt{q/h_p}\sim 0.013$  (Paardekooper et al. 2010) which means 
that the horseshoe region is resolved by about 12 grid cells in the radial direction. 
Also, we note that in calculating the torque experienced by the 
planet, we include the disc material located inside the Hill sphere of the planet.

\subsection{Total torque exerted on the planet}

As shown by Baruteau \& Lin (2010)  in the case of isothermal discs, turbulence can unsaturate the  vortensity-related part of the  corotation torque 
in the same way as an alpha viscosity does in a laminar disc.
In the following, we make the  assumption that thermal diffusion resulting from turbulence can also unsaturate 
the entropy-related part of the corotation torque and assume that this effect can be modelled 
by an entropy diffusion coefficient. This will be checked in the next section. \\
In the presence of turbulence, the torque exerted on a low-mass planet embedded in a turbulent disc can be considered as the 
sum of the differential Lindblad torque $\Gamma_L$, the corotation torque $\Gamma_{c}$ and the turbulent 
torque $\Gamma_{turb}$. In the adiabatic limit and for a smoothing parameter $b=0.4h_p$, 
the Lindblad torque is given by (Paardekooper et al. 2010):
\begin{equation}
\Gamma_L=(-2.5-1.7\beta+0.1\sigma)\frac{\Gamma_0}{\gamma_{ad}}
\label{eq:gammal}
\end{equation}
where $\Gamma_0$ is given by:
\begin{equation}
\Gamma_0=\left(\frac{q}{h_p}\right)^2\Sigma_pR_p^4\Omega_p^2
\end{equation}
In the previous equation, the subscript "p" 
indicates that quantities are evaluated at the orbital position of the planet $R_p$. \\
 Provided the surface density profile is not strongly 
affected by turbulence, we expect the barotropic part of the corotation torque to cancel for our disc model  in which
$\sigma=1.5$, since the latter scales 
with the vortensity gradient. In that case, the corotation torque consists only
 of its entropy-related part $\Gamma_{c,e}$ and the total torque $\Gamma$ exerted on the planet simplifies into:
\begin{equation}
\Gamma=\Gamma_L+\Gamma_{c,e}+\Gamma_{turb}
\end{equation}
 For diffusion timescales across the horseshoe region shorter than the horseshoe libration period 
$\tau_{lib}=8\pi R_p/3\Omega_px_s$ but longer than the horseshoe U-turn timescale 
$\tau_{U-turn}\sim h_p \tau_{lib}$ (Baruteau \& Masset 2008), 
the entropy-related horseshoe drag is close to its fully unsaturated value given (Paardekooper et al. 2010):
%\begin{equation}
%\Gamma_{hs,v}=1.1\frac{\Gamma_0}{\gamma_{ad}}\left(\frac{3}{2}-\sigma\right)
%\end{equation}
\begin{equation}
\Gamma_{hs,e}=\frac{\Gamma_0}{\gamma_{ad}}\left(\frac{7.9\xi}{\gamma_{ad}}\right)
\label{eq:gammahse}
\end{equation}
%\begin{equation}
%\Gamma_{c,lin,v}=0.7\frac{\Gamma_0}{\gamma_{ad}}\left(\frac{3}{2}-\sigma\right)
%\end{equation}  
while for diffusion timescales smaller than the U-turn timescale, $\Gamma_{c,e}$  decreases toward the entropy-related part of the linear 
corotation torque $\Gamma_{c,lin,e}$ with:
\begin{equation}
\Gamma_{c,lin,e}=\frac{\Gamma_0}{\gamma_{ad}}\left(2.2-\frac{1.4}{\gamma_{ad}}\right)\xi
\label{eq:gammacline}
\end{equation}
A model of the entropy-related corotation torque including its dependence on the diffusion timescale can be obtained 
by writing $\Gamma_{c,e}$ as a linear combination of $\Gamma_{hs,e}$ and $\Gamma_{c,lin,e}$. Paardekooper et al. (2011) 
have shown that $\Gamma_{c,e}$ can be indeed well approximated by:
%\begin{equation}
%\Gamma_{c,v}=F(p_\nu)G(p_\nu)\Gamma_{hs,v}+(1-K(p_\nu))\Gamma_{c,lin,v}
%\end{equation}
\begin{eqnarray}
\Gamma_{c,e} = & F(p_\nu)F(p_\chi)\sqrt{G(p_\nu)G(p_\chi)}\Gamma_{hs,e}\\+
\label{eq:gammace}
&\sqrt{(1-K(p_\nu))(1-K(p_\chi)}\Gamma_{c,lin,e}\nonumber
\end{eqnarray}
where $p_\nu$ is the saturation parameter related to viscosity and is given by:
\begin{equation}
p_\nu=\frac{2}{3}\sqrt{\frac{R_p^2\Omega_p x_s^3}{2\pi \nu}}
\end{equation} 
and $p_\chi$ is the saturation parameter related to thermal diffusion with:
\begin{equation}
p\chi=\frac{2}{3}\sqrt{\frac{R_p^2\Omega_p x_s^3}{2\pi \kappa}}
\end{equation}
where $\nu=\alpha c_s H$ with $\alpha$ given by Eq. \ref{eq:fitalphav} and where $\kappa$ is related to the amplitude of 
turbulence by Eq. \ref{eq:fitkappa}.
In  Eq. $28$, the function $F(p)$ of the parameter $p$ is defined by:
\begin{equation}
F_p=\frac{1}{1+(p/1.3)^2}
\end{equation}
while the functions $G(p)$ and $K(p)$ are given by:
\begin{equation}
G(p)=
\begin{cases}
\frac{16}{25}\left(\frac{45\pi}{8}\right)^{3/4}p^{3/2}, & \text{if}\; p < \sqrt{8/(45\pi)} \\
1-\frac{9}{25}\left(\frac{8}{45\pi}\right)^{4/3}p^{-8/3}, & \text{if}\; p\ge \sqrt{8/(45\pi)} 
\end{cases}
\end{equation}

and:
\begin{equation}
K(p)
\begin{cases}
\frac{16}{25}\left(\frac{45\pi}{28}\right)^{3/4}p^{3/2}, & \text{if}\; p < \sqrt{28/(45\pi)} \\
1-\frac{9}{25}\left(\frac{28}{45\pi}\right)^{4/3}p^{-8/3}, & \text{if}\; p\ge \sqrt{28/(45\pi)} 

\end{cases}
\end{equation}
Regarding the turbulent torque $\Gamma_{turb}$; its amplitude is given by (Nelson 2005; 
Baruteau \& Lin 2010):
\begin{equation}
|\Gamma_{turb}|=\sigma_{turb}\left(\frac{t}{\tau_c}\right)^{-1/2}
\label{eq:gammaturb}
\end{equation}

where $\sigma_{turb}\sim 2.4\times 10^2 \Sigma_p q \gamma R_p^4 \Omega_p^2$ (Baruteau \& Lin 2010) 
is the mean deviation of the turbulent torque distribution and 
$\tau_c\sim 0.5\; T_{orb}$ is the correlation time of the turbulence. Using the previous 
equation leads to an estimation of  the convergence time $t_{conv}$ of the simulations 
over which  $|\Gamma_{turb}|$ is a 
small fraction 
$f$ of $|\Gamma_L+\Gamma_{c,e}|$.  Using Eq. \ref{eq:gammaturb}, the convergence time is given by:
\begin{equation}
t_{conv}\sim \tau_c f^{-2}\left(\frac{\sigma_{turb}}{\Gamma_L+\Gamma_{c,e}}\right)^2
\end{equation}
Previous 
equation shows that the maximum convergent time of the turbulent runs is reduced in the case where both the entropy-related corotation  torque and 
the differential Lindblad torque have the same sign. In order to minimize $t_{conv}$ and given that the differential Lindblad torque is negative, 
we employed a positive initial entropy gradient so that the entropy-related corotation torque is negative as well.
Using the expression for $\Gamma_L$ given by Eq. \ref{eq:gammal} and taking the fully unsaturated value  for the
entropy-related corotation torque given by Eq. \ref{eq:gammahse}, we estimate the  convergence time 
to be approximately  given by:
\begin{equation}
t_{conv}\sim 0.2\left(\frac{\gamma}{q}\right)^2T_{orb},
\label{eq:tconv}
\end{equation} 
 where we assumed $f=0.1$. For $q=5\times 10^{-6}$ and $\gamma=10^{-4}$, this gives $t_{conv}\sim 100\; T_{orb}$   
%A better approximation for $t_{conv}$ may be obtained by considering the full expression for the entropy-related corotation 
%torque. In that case, we find that the convergence time is increased by $\sim 50\; \%$ for $\gamma=2\times 10^{-5}$ while it 
%is increased by $\sim 20 \; \%$ for $\gamma=10^{-4}$.}
 Over timescales typical to those covered by our simulations 
($\sim 1500$ $T_{orb}$ ) and 
provided that there is no significant coupling between the Lindlad torque and the turbulent torque, we therefore 
expect the steady-state torque experienced by the planet to be approximately given by:
\begin{equation}
\Gamma\sim \Gamma_L+\Gamma_{c,e}
\label{eq:totaltorque}
\end{equation}
where $\Gamma_L$ and $\Gamma_{c,e}$ are respectively given by Eqs. \ref{eq:gammal} and \ref{eq:gammace}. We emphasize that the previous 
equation is valid in the case where the effect of turbulent thermal diffusion on the desaturation of 
the entropy-related corotation torque is similar to that corresponding to a laminar entropy diffusion coefficient. The aim of the 
following section is to check this assumption.

\subsection{Time evolution of the running time-averaged torque}
\begin{figure*}
\centering
\includegraphics[width=0.49\textwidth]{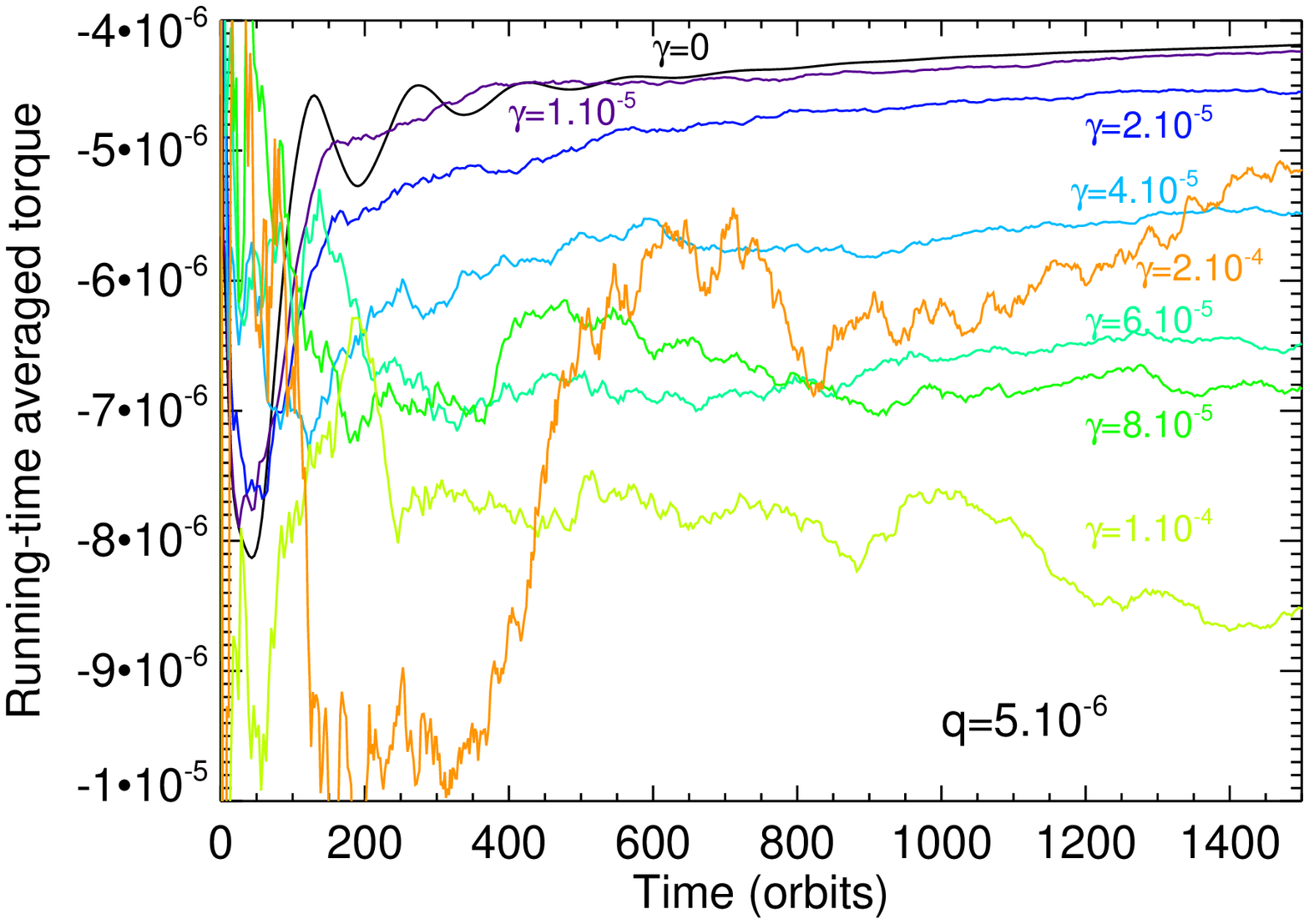}
\includegraphics[width=0.49\textwidth]{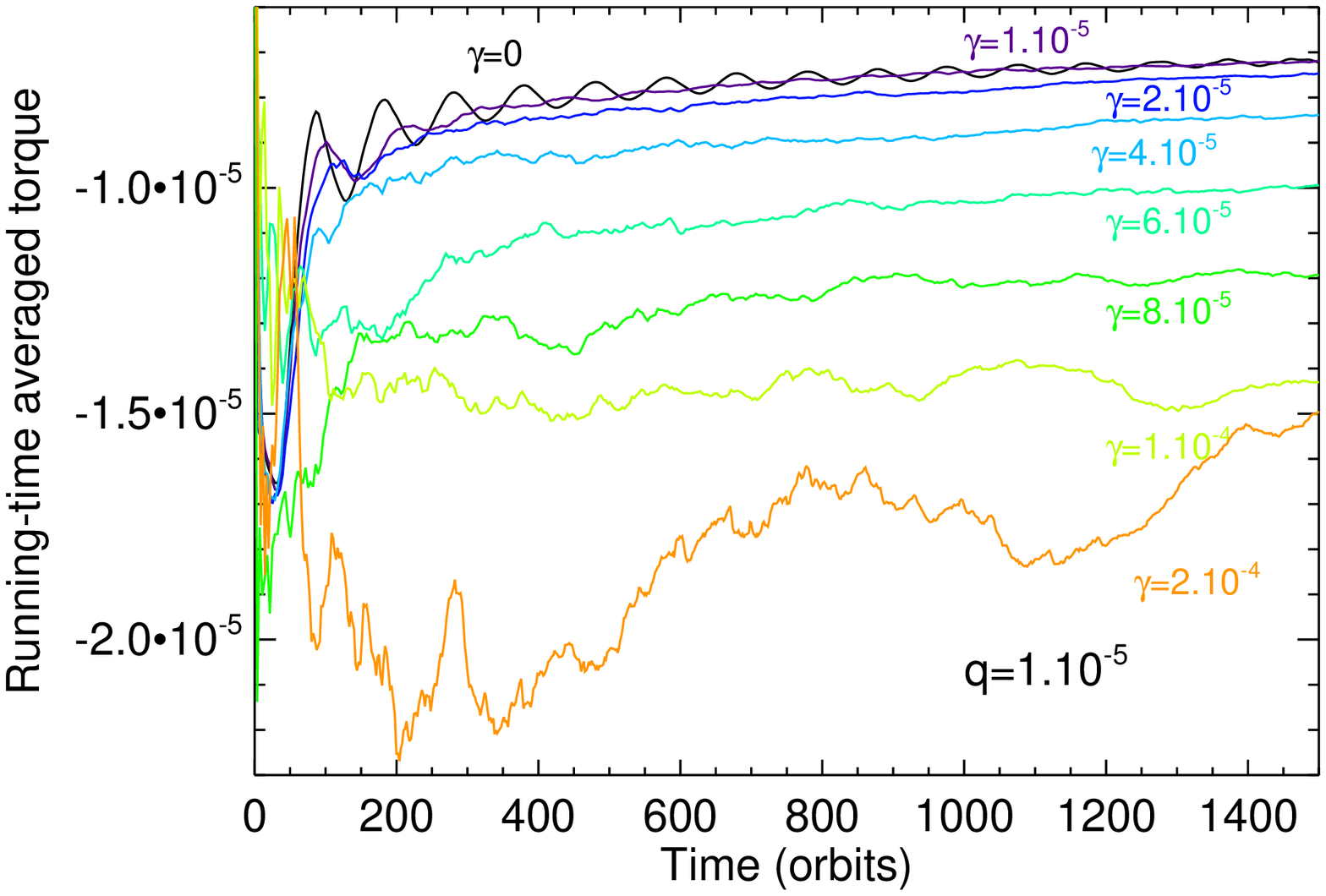}
\caption{{\it Left panel:} Time evolution of the running time-averaged specific torque for the different values of $\gamma$ we considered and for $q=5\times 10^{-6}$. {\it Right panel:} same but for $q=10^{-5}$.}
\label{fig:torque}
\end{figure*}

For $q=5\times 10^{-6}$, the time evolution of the running time-averaged torque over $1500$ orbits for turbulent 
runs with $\gamma$ in the range $10^{-5}-2\times 10^{-4}$ is displayed in the left panel of 
Fig. \ref{fig:torque}.  From Eq. \ref{eq:tconv}, we expect that the running time-averaged torques have 
converged over this timescale  except for the run with $\gamma=2\times 10^{-4}$ in which the temperature 
profile is found to be continuously altered by turbulence (see Sect. \ref{sec:analytics}). Also shown is the result of an inviscid simulation with $\gamma=0$ in which the torque
saturates toward the differential Lindblad torque $\Gamma_L$ . In that case, $\Gamma_L$ is found to slowly increase 
by $\sim 2\%$ between $1000$ and $1500$ orbits due to the continuous formation of a gap around the planet. We note 
that such a  process was also observed in the isothermal turbulent simulations of Baruteau \& Lin (2010).\\
In turbulent runs, the running time-averaged torque typically reaches a quasi-stationary value after $\sim 200$ orbits for moderate 
values of $\gamma$, which is relatively consistent with the estimation of the convergence time given by 
Eq. \ref{eq:tconv}. Up to $\gamma=10^{-4}$, we see that this stationary value decreases as $\gamma$ increases whereas for 
higher values of $\gamma$,  it tends to increase with $\gamma$. As the differential Lindblad torque 
should not be significantly altered by turbulence (Baruteau \& Lin 2010, Baruteau et al. 2011) and since the 
barotropic part of the corotation torque is expected to cancel for the chosen surface density profile ($\sigma=1.5$), 
this seems to suggest that turbulence  does unsaturate the entropy-related corotation torque. In that case, we indeed expect 
that as $\gamma$ increases, the torque experienced by the planet decreases from the differential Lindblad torque down 
to value close to that corresponding to the  sum of the differential Lindblad torque plus the unsaturated entropy-related horseshoe drag. 
Such a value is reached when the thermal diffusion 
timescale across the horseshoe region $\tau_d=x_s^2/\kappa$ is approximately equal to half the libration timescale $\tau_{lib}=8\pi R_p/(3\Omega_p x_s)$. Using Eq. \ref{eq:fitkappa}, we estimate the unsaturated entropy-related horseshoe drag 
to be obtained for:
\begin{equation}
\gamma\sim 0.07\left(\frac{q}{h_p}\right)^{3/4}
\label{eq:gammacr}
\end{equation}
Using Eq. \ref{eq:fitalphav}, this corresponds to:
\begin{equation}
\alpha \sim  0.6q^{3/2}h_p^{-7/2}
\end{equation}

For $q=5\times 10^{-6}$, this gives  $\gamma\sim 1.1\times 10^{-4}$ ($\alpha\sim 1.6\times 10^{-3}$) which is close to the value observed in the 
simulations. For higher values of $\gamma$, the diffusion timescale across the horseshoe region can approach 
the U-turn timescale $\tau_{U-turn}\sim h_p \tau_{lib}$ (Baruteau \& Masset 2008) and the amplitude of the entropy-related 
corotation torque starts to decrease toward a value corresponding to that predicted by the linear theory.\\
The results for $q=10^{-5}$ are shown in the right panel of Fig. \ref{fig:torque} which presents the time evolution of the 
running time-averaged torque for $\gamma \le 2\times 10^{-4}$. For $\gamma=0$, it appears that the torque  oscillates 
around a value corresponding to the differential Lindblad torque.  As discussed in Baruteau \& Lin (2010), these 
oscillations are due to the formation of vortices flowing along the edge of the gap created by the planet 
(Li et al. 2005). However, the running time-averaged torque does converge toward the 
differential Lindblad torque but compared with the case $q=5\times 10^{-6}$,  the torque saturates much more 
slowly. 
%  A similar behaviour was discussed in Masset \& Casoli (2010) who 
%argued that the torque experienced by the planet can indeed significantly oscillates before saturating in the case where most of the vortensity 
%created in the vicinity of the planet executes horseshoe U-turns rather than cirulating. \\
For $\gamma \ne 0$, we estimate using Eq. \ref{eq:gammacr} that the value
of $\gamma$ from which the entropy-related horseshoe drag starts to decrease toward its linear value is $\gamma\sim 1.8\times 10^{-4}$. 
This is in decent agreement with the results of the simulations for $q=10^{-5}$ which show, in the limit of 
$\gamma \le 2\times 10^{-4}$, a  tendency for the torque amplitude to increase with $\gamma$. Again, this enlightens the process of desaturation of the entropy-related corotation torque 
by turbulence.

%\subsection{Comparison with analytics}
\subsection{Comparison with laminar viscous disc models}
\label{sec:analytics}
\begin{figure*}
\centering
\includegraphics[width=0.49\textwidth]{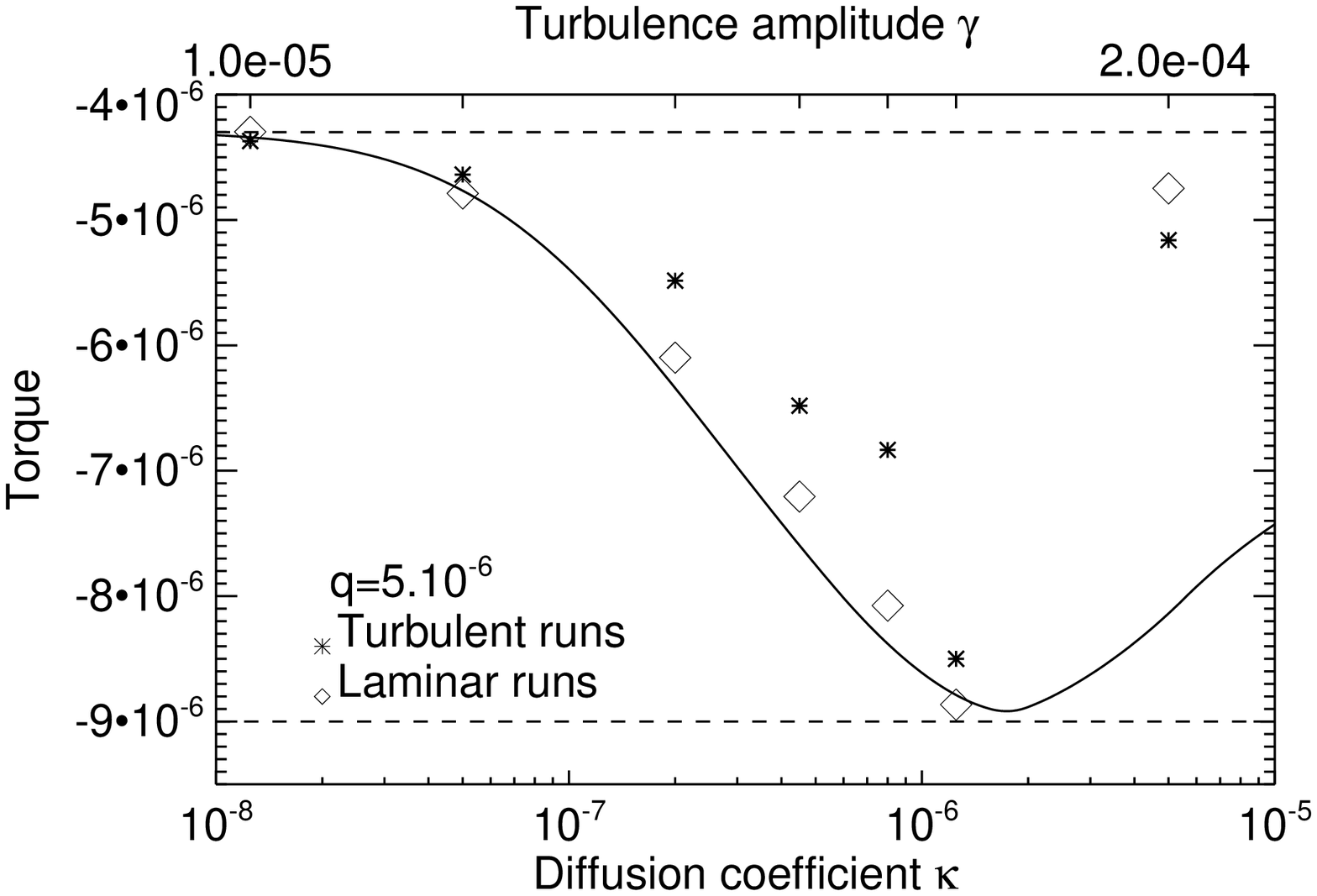}
\includegraphics[width=0.49\textwidth]{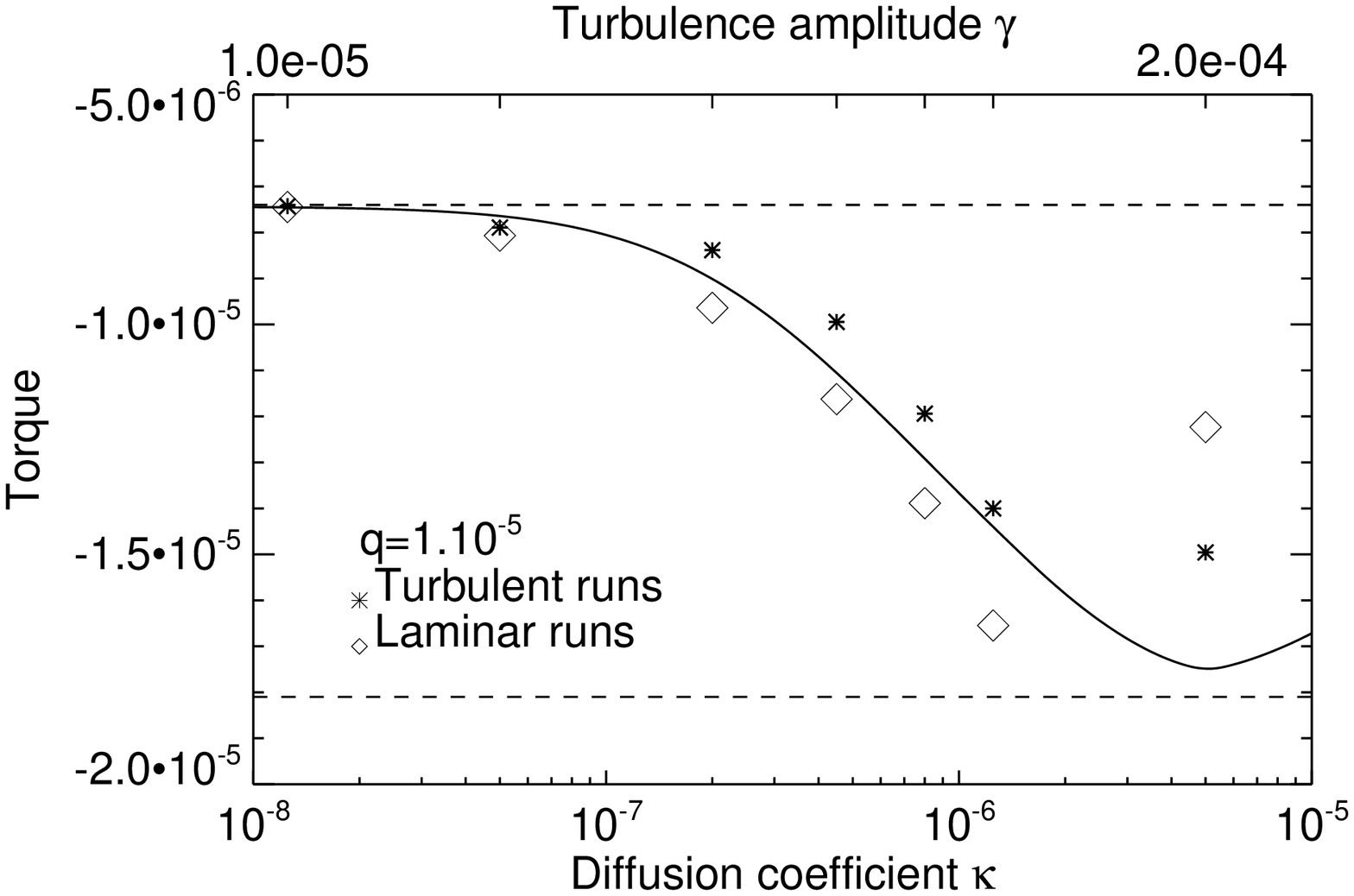}
\caption{ Running time-averaged specific torque (stars) in turbulent runs as a function of the turbulence amplitude $\gamma$ (top axis) and as a 
function of the effective entropy diffusion coefficient $\kappa$ (bottom axis) computed using Eq. \ref{eq:fitkappa}. Diamonds
 correspond to the results of laminar simulations with values for the viscous and thermal diffusion coefficients equal 
to those effective in turbulent runs. The initial surface density and entropy profiles in laminar runs correspond to 
the profiles in turbulent runs time averaged over $1500$ $T_{orb}$. The solid line depicts the analytic estimation of Paardekooper et al. (2011) 
expected for the torque as a function of the thermal diffusivity $\kappa$, for a kinematic viscosity $\nu=S_c^{-1}P_r \kappa= 
1.2\kappa$ and using the initial surface density and entropy profiles in turbulent runs. The upper and lower dashed lines 
depict the values of the differential Lindblad torque and the fully unsaturated torque respectively.}
\label{fig:rta}
\end{figure*}

\begin{figure}
\centering
\includegraphics[width=0.95\columnwidth]{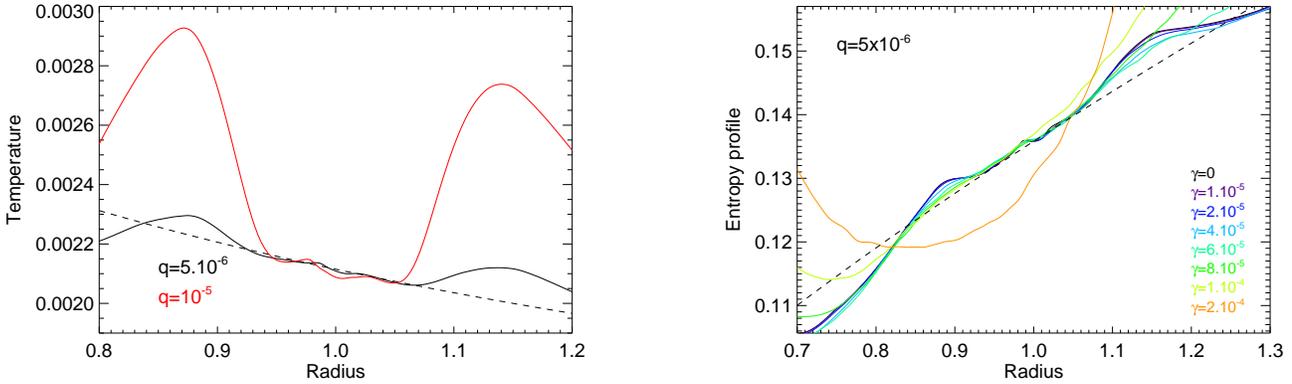}
\caption{Temperature profile at $t=1000$ $T_{orb}$ for $q=5\times 10^{-6}$ black line and $q=10^{-5}$ (red line).  The 
dashed line corresponds to the initial temperature profile.}
\label{fig:tinviscid}
\end{figure}

\begin{figure}
\centering
\includegraphics[width=0.95\columnwidth]{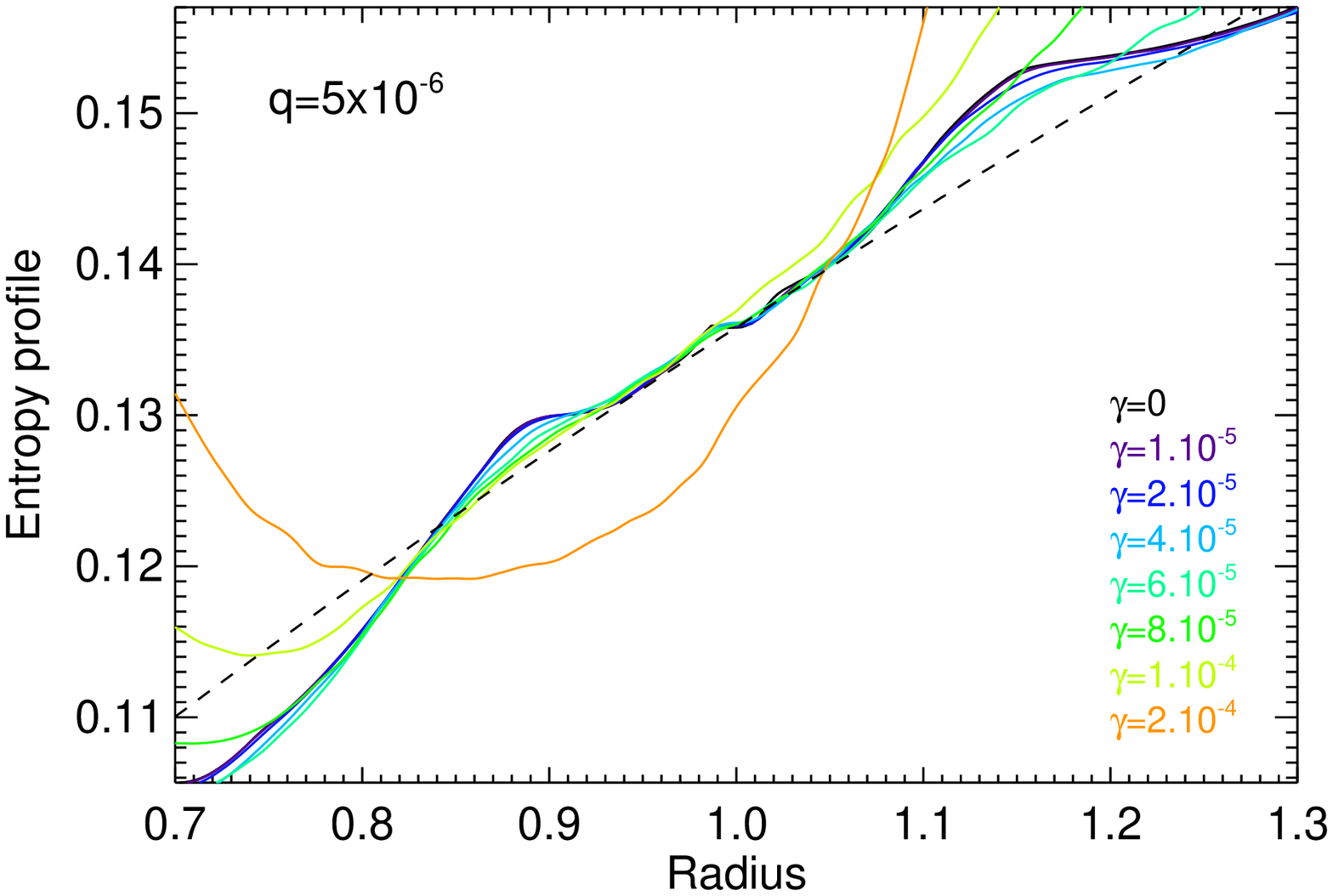}
\includegraphics[width=0.95\columnwidth]{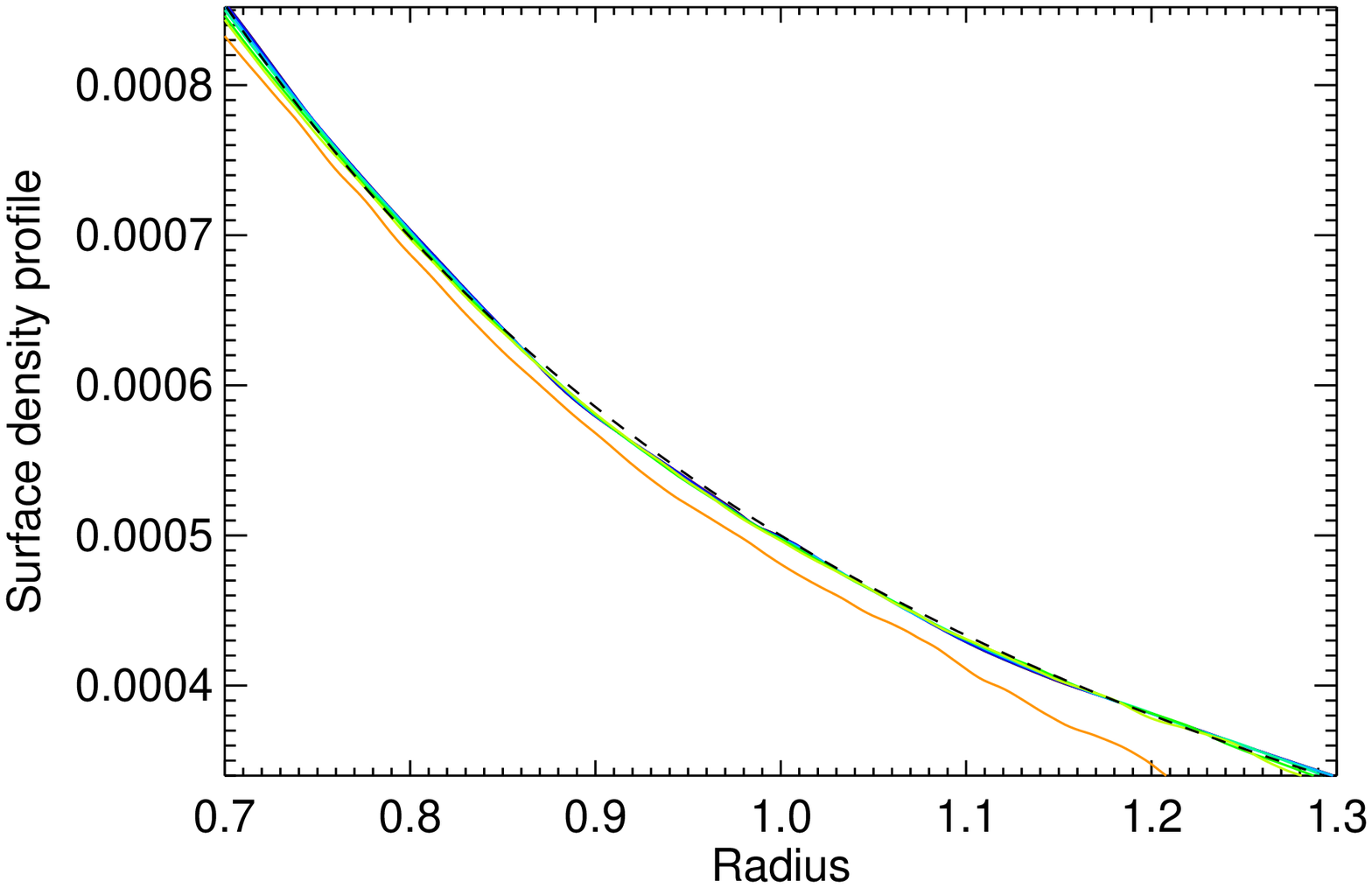}
\caption{{\it Upper panel:} Entropy profile in turbulent runs with $q=5\times 10^{-6}$ and time averaged over $1500$ $T_{orb}$. 
 For $\gamma=2\times 10^{-4}$, the entropy profile was offset by a factor of $2$ for clarity.  The dashed line 
corresponds to the initial entropy profile. {\it Lower panel:} Same but for the surface density profile.}
\label{fig:tprofile}
\end{figure}

In this section, we compare the values for the running time-averaged torques deduced from the turbulent runs with the torque 
formulae for non-isothermal Type I migration  derived by Paardekooper et al. (2011)  and with laminar 
simulations with constant values for the kinematic viscosity and thermal diffusivity. In Fig. \ref{fig:rta} are 
plotted, for $q=5\times 10^{-6}$ (left panel) and $q=10^{-5}$ (right panel), the running time-averaged torques computed at $t=1500$ $T_{orb}$ as a function of 
the estimated entropy turbulent diffusion coefficient $\kappa$. The top axis gives the corresponding value for the  $\gamma$ parameter which is related to $\kappa$ 
by the relationship given in Eq. \ref{eq:fitkappa}. The solid line in Fig. \ref{fig:rta} depicts the approximation for 
the total torque given  by Eq. \ref{eq:totaltorque} where we remind  the reader that $\Gamma_L$ is the Lindblad torque and $\Gamma_{c,e}$ the  
entropy-related corotation torque given by Eq. \ref{eq:gammace}. In this latter relation, we computed the saturation 
parameter related to viscosity $p_\nu$ by setting $\nu=S_c^{-1} P_r\kappa\sim 1.2 \kappa$. Moreover, we used the results from 
the run  with 
$\gamma=0$ and in which the corotation torque is expected to saturate to determine $\Gamma_L$. From this simulation we measure $\Gamma_L\sim 1.54\Gamma_0$
for $q=5\times 10^{-6}$ and $\Gamma_L \sim 1.34 \Gamma_0$ for $q=10^{-5}$ which differs from the estimation for the Lindblad torque given in Eq. \ref{eq:gammal} by $\sim 15 \%$ and 
$\sim 25 \%$ respectively. We interpret this 
slight discrepancy as being due to a change of the temperature profile in the vicinity of the planet. 
This is exemplified in Fig. \ref{fig:tinviscid} which displays for the two values of $q$ we considered the temperature profile in the case where $\gamma=0$.  
It reveals that the temperature tends to increase at distances 
$\pm 2H$ from the planet, corresponding to a local increase of the disc aspect ratio and resulting consequently in a decrease of the 
Lindblad torque. \\ 
For the run with $\gamma=0$, the fully unsaturated horseshoe drag was evaluated to be $\Gamma_{hs,e}\sim 1.82 \Gamma_0$ for $q=5\times 10^{-6}$ and $\Gamma_{hs,e}=1.92\Gamma_0$ for $q=10^{-5}$ which are values close to the estimation given by Eq. \ref{eq:gammahse} which predicts $\Gamma_{hs,e}\sim 1.7 \Gamma_0$. Regarding the linear corotation torque, it can be  crudely estimated 
by  subtracting the differential Lindblad torque from the total torque at $t\sim 2$ $T_{orb}$ ( that is,  
$t\sim 0.5\tau_{U-turn}$; Baruteau \& Masset 2008;  Paardekooper \& Papaloizou 2008; Casoli \& Masset 2010) and this 
gives $\Gamma_{c,lin,e}\sim 0.71 \Gamma_0$ for $q=5\times 10^{-6}$ and $\Gamma_{c,lin,e}\sim 0.86\Gamma_0$ for $q=10^{-5}$ 
while using Eq. \ref{eq:gammacline} leads to $\Gamma_{c,lin,e}\sim 0.46 \Gamma_0$.\\
For modest values of $\gamma$, we see that good agreement is obtained between the results of the turbulent runs 
and the analytic formulae of  
Paardekooper et al. (2011). The basic reason for the differences at higher values of $\gamma$ possibly arises  
from the evolution of the disc temperature profile. This is illustrated in Fig. \ref{fig:tprofile} which shows for 
$q=5\times 10^{-6}$ the entropy and surface density profiles time-averaged over $1500$ $T_{orb}$. We note that in Fig. 
\ref{fig:tprofile} the entropy profile corresponding to $\gamma=2\times 10^{-4}$ was offset by a factor of $2$ for clarity. Although the density remains 
almost unchanged, the entropy profile can be significantly affected in the vicinity of the planet for $\gamma \ge 10^{-4}$ and this should in principle affect the value for the total torque. 
For $\gamma=10^{-4}$, we find that with respect to an inviscid run, the entropy gradient is  increased by $\sim 20 \%$ at the location of the planet, which should increase the amplitude of the (negative) entropy-related corotation torque by the same amount. The corresponding change 
on the slope of the temperature profile is estimated to lead to a $\sim 13 \%$ decrease in the amplitude of the (negative) differential Lindblad torque. Interestingly, such changes in the values for the corotation and the differential Lindblad torques 
are found to almost cancel each other which suggests that in that case the modification of the 
temperature by turbulence has little impact on the total torque experienced by the planet.  
In the calculation with $\gamma=2\times 10^{-4}$, the  surface density at $R=1$ in only slightly reduced by $\sim 3\%$ 
with respect to the initial profile  but the temperature is found to be higher by a 
factor of $\sim 2$, with the consequence that the torque exerted on the planet is weaker by the same factor compared to a 
laminar run. Moreover, we quantify the corotation torque and the differential Lindblad torque 
to be altered by $\sim 57 \%$  and $\sim 65 \%$ respectively due to the modification of the temperature slope at the 
position of the planet. From these values, we estimate that for the run with $\gamma=2\times 10^{-4}$ 
and $q=5\times10^{-6}$, the 
specific torque exerted on the planet would be $\sim 10^{-5}$ in the absence of modification of the temperature 
profile by turbulence, which is somewhat larger than the value predicted by Eq. \ref{eq:totaltorque}. We note 
that differences between the time-averaged turbulent torques and the analytical estimations may be also related to 
uncertainties in Eqs. \ref{eq:fitalphav} and \ref{eq:fitkappa} which relate the value of $\gamma$ to those of $\alpha$ and 
$\kappa$. \\ 

%\subsection{Comparison to laminar runs}
%\subsection{Comparison with laminar viscous disc models}

\begin{figure*}
\centering
\includegraphics[width=0.325\textwidth]{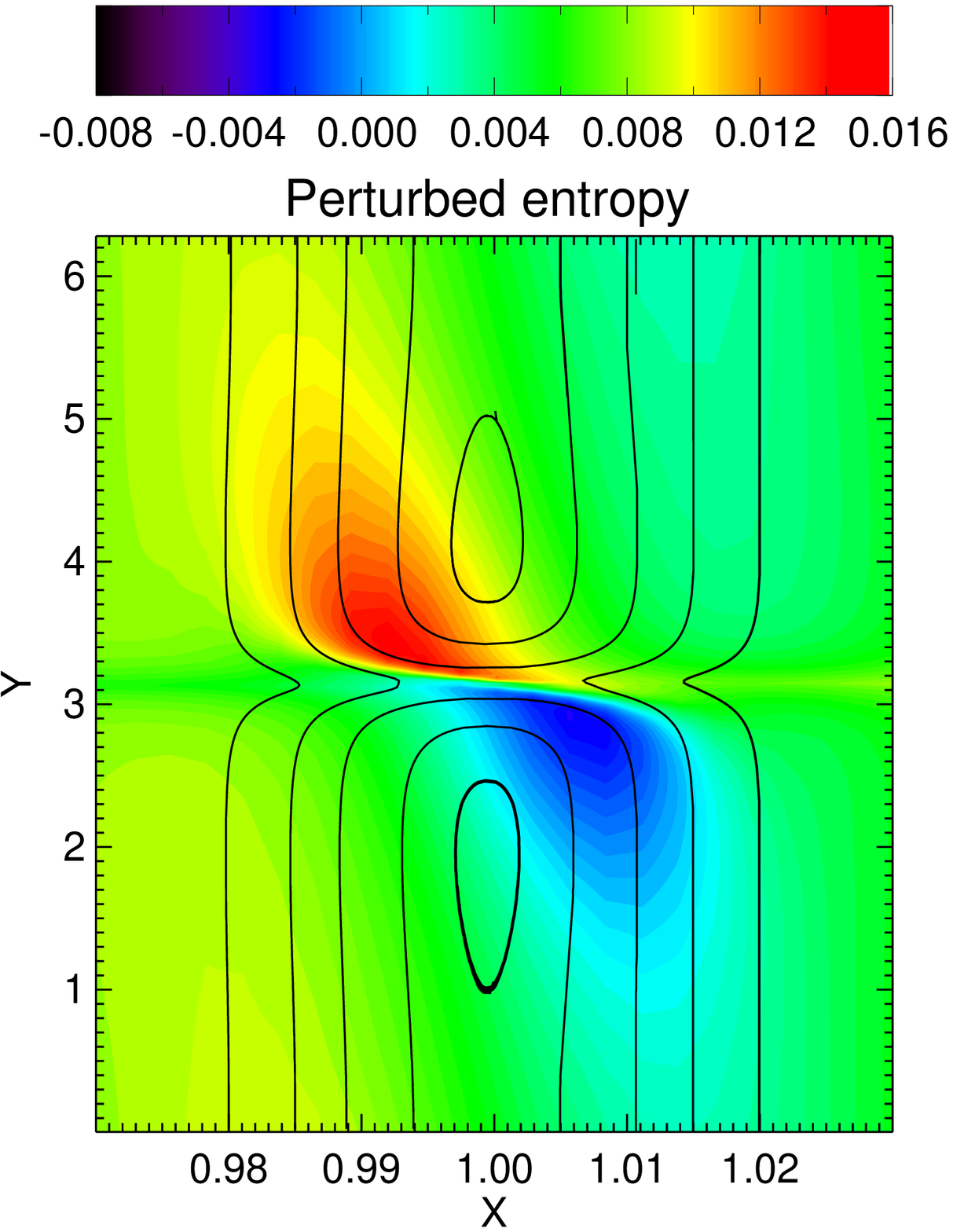}
\includegraphics[width=0.325\textwidth]{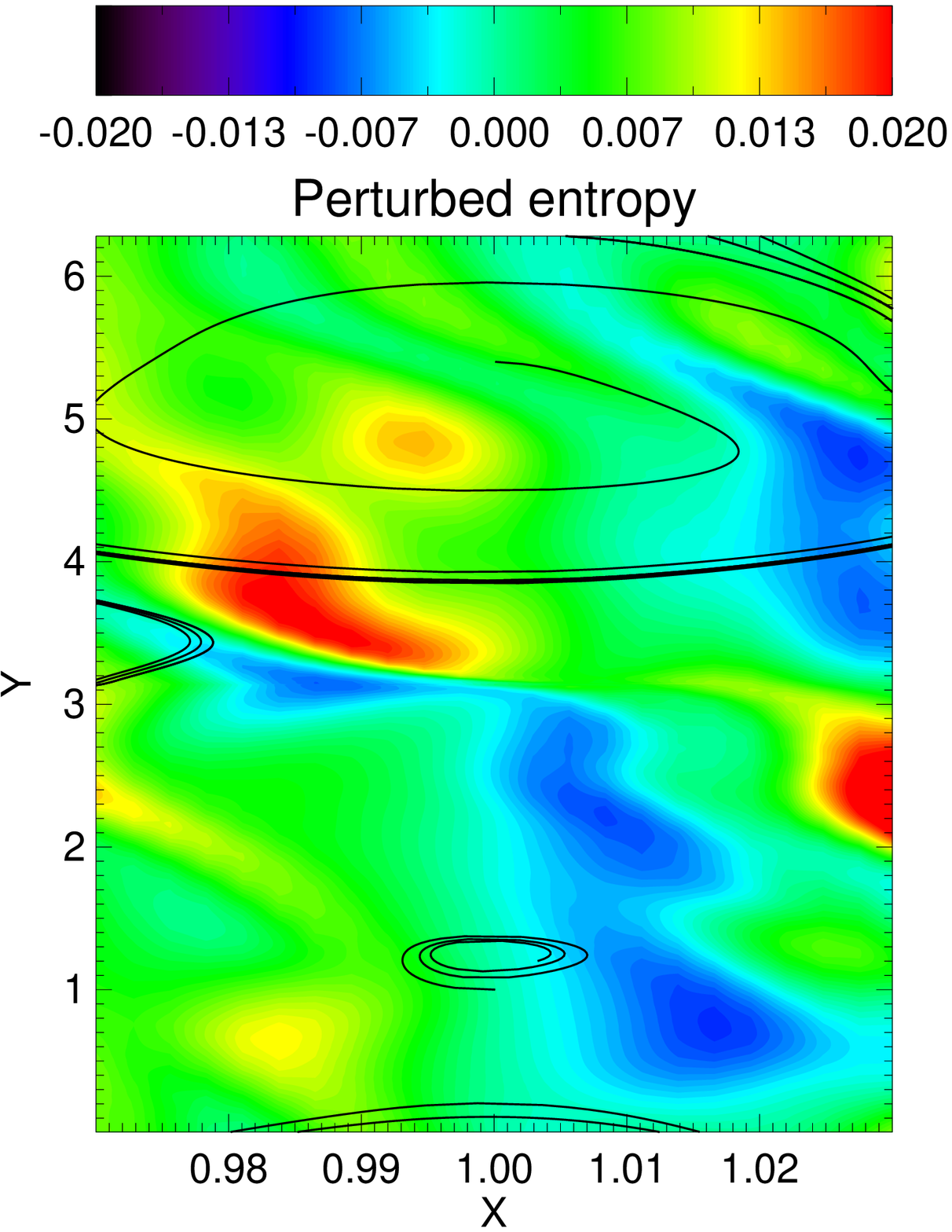}
\includegraphics[width=0.325\textwidth]{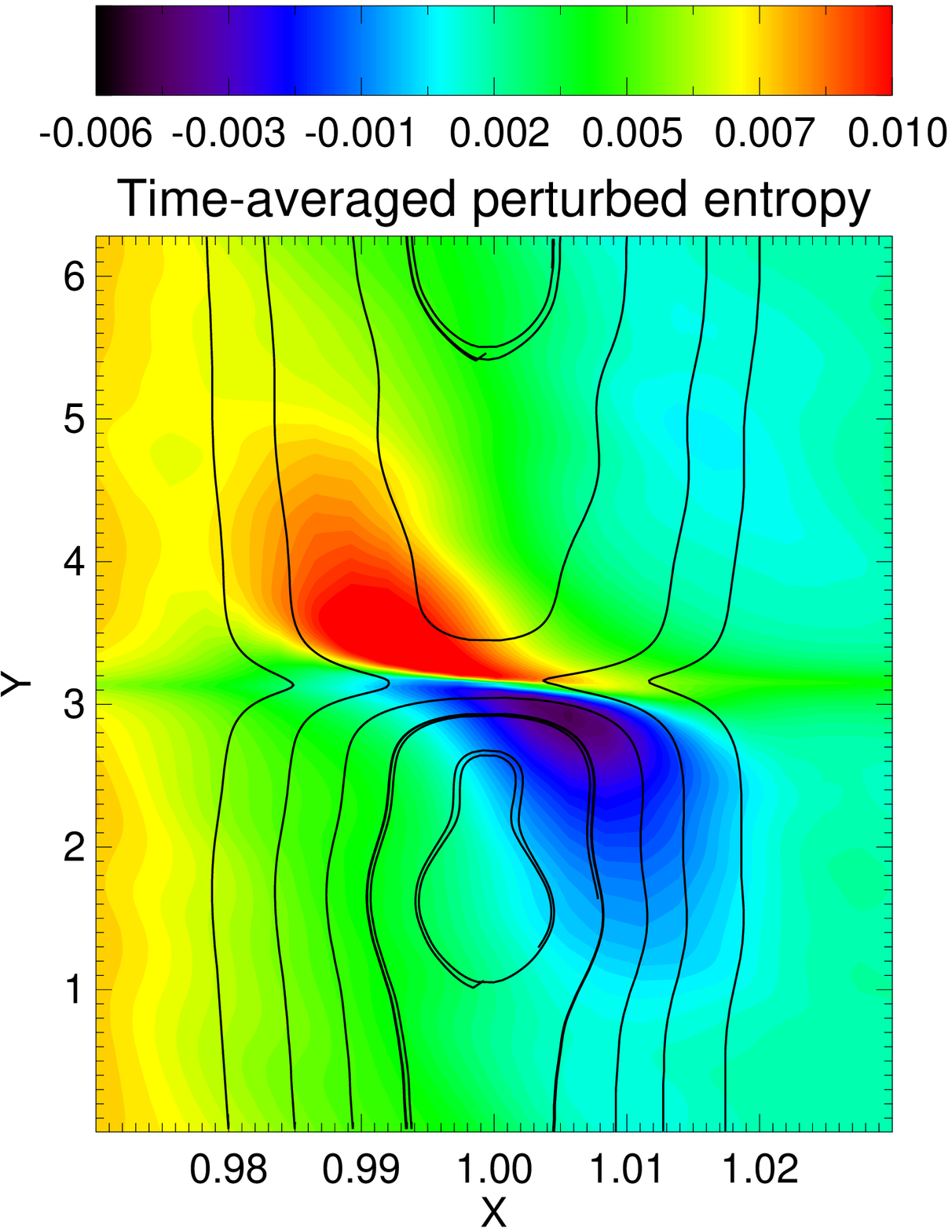}
\caption{{\it Left panel:} Snapshot of the disc perturbed entropy at $t=800\;T_{orb}$ for a laminar run with  
$q=5\times 10^{-6}$ and with constant thermal and viscous diffusion coefficents with values $\kappa=4.5\times 10^{-7}$ and $\nu=5.4\times 10^{-7}$ respectively. Streamlines are overplotted as black lines. {\it Middle panel:} same but for a turbulent 
run with $\gamma=6\times 10^{-5}$. {\it Right panel:}  Contour plot of the perturbed entropy for the same turbulent 
run but time-averaged 
 between $750$ and $850$ orbits and over $6$ different realizations. Averaged streamlines are overplotted as black lines. }
\label{fig:2dplot}
\end{figure*}

We now turn to the question of how the torque dependence upon turbulence strength in turbulent simulations 
compares with the torque dependence upon kinematic viscosity and thermal diffusivity in laminar runs. To investigate 
this issue, we have performed an additional suite of laminar simulations in which we vary the value for 
the thermal diffusivity while  the kinematic viscosity is set to $\nu= S_c^{-1}P_r\kappa\sim 1.2 \kappa$. The values for the thermal
diffusivity are chosen in such a way as to correspond to the effective values for thermal diffusivity in 
turbulent runs and estimated using Eq. \ref{eq:fitkappa}. In order to take into account the alteration of the 
temperature profile at high levels of turbulence, the inital surface 
density and temperature profiles in these laminar simulations consist of the surface density and temperature profiles 
in turbulent runs time-averaged over $1500$ orbits. The laminar torques typically reach a quasi-stationary value by $500$ 
 $T_{orb}$ and they are directly compared to the running time-averaged turbulent torques at $t=1500$ $T_{orb}$ in Fig. \ref{fig:rta}.
 Again, for the two values of $q$ we consider, relatively good agreement is obtained between the laminar and turbulent runs, even for high values of the thermal diffusivity for which the temperature profile can be strongly modified in turbulent runs. One can notice 
however a tendency for the turbulent torques amplitudes to be slightly smaller than the laminar torques, which would 
suggest that for similar values of the thermal and viscous diffusion coefficients, the corotation torque is slightly more saturated in turbulents discs than in laminar viscous disc models.  We note that such a trend was not observed in previous studies 
of  the vorticity-related corotation torque in turbulent discs (Baruteau \& Lin 2010; Baruteau et al. 2011) which suggested 
that the turbulent torque experienced by a protoplanet embedded in a isothermal turbulent disc is in decent agreement with its counterpart in a laminar 
viscous disc model.  This discrepancy may arise because in non-isothermal discs, the 
entropy-related corotation torque is associated with singular production of vortensity on downstream separatrices and 
we may expect that compared with a turbulent disc, such a process is slightly more efficient in a laminar viscous disc model. 
%For $q=5\times 10^{-6}$, one can see that very good agreement is obtained between the laminar and turbulent runs, even 
%for high values of the thermal diffusivity for which the temperature profile can be strongly modified in turbulent runs. 
%Differences between the laminar and turbulent torques are more pronounced for $q=10^{-5}$. This probably arises 
%not only because saturation of the corotation torque proceeds much more slowly  for low values of the 
%thermal diffusivity (see Sect. \ref{sec:analytics}) due to the fomation 
%of vortices in the vicinity of horseshoe U-turns but also because gap opening 
%occurs faster  in that case, which causes the value of the saturated torque to significantly increase with time. 
The fair agreement between turbulent and laminar torques reveals not only that the total torque experienced by a protoplanet 
 can be decomposed into a laminar torque plus a stochastic torque, which was already suggested by Baruteau \& Lin (2010)
 in the case of isothermal disc models, but also that the entropy-related horseshoe drag takes similar values in 
 turbulent and laminar disc models. This suggests that the structure of the horseshoe region 
in turbulent discs should be similar  in time average to that in laminar disc models, which can be confirmed by 
comparing  contour plots of the perturbed entropy for a turbulent run to those 
corresponding to a laminar 
calculation with similar values for the viscous and thermal diffusion coefficients. In Fig. 
\ref{fig:2dplot} we present the perturbed entropy field  for a laminar 
simulation (left panel) with $\kappa=4.5\times 10^{-7}$ and $\nu=5.4\times 10^{-7}$ and for a turbulent 
run (middle panel) with $\gamma=6\times 10^{-5}$ (which correspond to effective values $\nu\sim 5.4\times 10^{-7}$ and $\kappa \sim 4.5\times 10^{-7}$) at $t=800\;T_{orb}$. 
Overplotted are a few streamlines which are indicated with black lines and which were computed using the same initial 
coordinates in the $R-\phi$ plane for the laminar and the turbulent runs. 
 The right panel displays, for the turbulent run, the corresponding entropy field
and  streamlines averaged between $750$ and $850$ orbits (using field outputs produced every planet orbit) and over $6$ different realizations. Interestingly, one can see that 
the averaged streamlines in the turbulent run have very similar shape to the streamlines in the laminar
simulation, which is in agreement with the results of Baruteau \& Lin (2010) in the case of isothermal discs. Here, 
since the initial entropy 
gradient is positive, dynamics in the horseshoe region yields a positive entropy perturbation at the  inward downstream 
separatrix ($\phi>\phi_p$; $R<R_p$) whereas a negative entropy perturbation is created at the  outward downstream separatrix ($\phi<\phi_p$; $R>R_p$). 
It is clear that these entropy perturbations for the turbulent run are very similar to those corresponding to 
the laminar calculation, which 
confirms  that the thermal diffusion coefficients are indeed close to each other in both 
cases.

\section{Discussion and conclusion}
In this paper, we have presented the results of 2D hydrodynamical simulations which study the interaction 
of an embedded low-mass planet with a turbulent 
non-isothermal disc. 
The aim is to investigate the role of turbulence on Type I migration in non-isothermal discs, with 
particular emphasis on the process of desaturation of the entropy-related corotation torque 
by turbulence. \\
We employ the turbulence model of Laughlin et al. (2004) in which a turbulent potential 
corresponding to the superposition of multiple wave-like modes is applied to the disc. This 
turbulence potential  had been shown previously to reproduce the main statistical properties of MHD turbulence
  in isothermal discs (Baruteau \& Lin 2010) . In non-isothermal discs, 
turbulent density and entropy fluctuations generate transport of momentum and heat inside the 
disc and we show that both the associated viscous and thermal diffusion coefficients scale as $\gamma^2$, 
where $\gamma$ refers to the amplitude of the turbulent forcing. This results in a turbulent 
Prandtl number $P_r=\alpha_R c_s H/\kappa$ where $\alpha_R$ is the Reynolds stress parameter  which is almost constant with $\gamma$ and we estimate  
 for the first time that $P_r \sim 0.3$ in non-isothermal turbulent discs.  We caution, however, that such a value for the turbulent Prandtl number should be checked against 3D 
MHD, non-isothermal simulations in which turbulence is generated by the MRI. \\
By investigating the evolution of the running time-averaged torque experienced by a protoplanet for 
different values of $\gamma$, we show that turbulence can unsaturate the entropy-related corotation 
torque.  For similar values for the viscous and thermal diffusion coefficients, we find that the formulae of Paardekooper 
et al. (2011) for non-isothermal Type I migration in laminar disc models reproduce quite satisfactorily the time-averaged torques 
deduced from our turbulent runs,  provided that the typical amplitude of turbulent fluctuations is not too large 
so that turbulent transport of heat and momentum do not significantly modify the background density and 
temperature profiles. For high values 
of $\gamma$  corresponding to  values for the dimensionless $\alpha$ viscosity of Shakura \& Sunyaev (1973) $\alpha \gtrsim 10^{-3}$, we indeed observe discrepancies between the analytical estimate and the results of simulations which, to 
a large extent, are related to the alteration of the temperature profile by turbulence.  
We find also that the amplitudes of the running time-averaged 
turbulent torques  are slightly smaller than the torques amplitudes deduced from laminar simulations with initial density and temperature 
profiles corresponding to the time averaged profiles resulting from turbulent runs. This suggests 
that the entropy-related horseshoe drag in presence of turbulence is slightly more saturated compared 
to an equivalent laminar disc model with similar vortensity and entropy diffusion coefficients. \\
Although these results were obtained from a disc model in which the initial entropy gradient is 
 positive, similar conclusions  should be drawn in the case of a negative initial entropy gradient.
 Since in that case
the amplitude of the total torque is only a small fraction of the typical amplitude of the stochastic torque, 
turbulent runs would require a much longer convergence time than the simulations presented here. 
For a negative initial entropy gradient, the 
entropy-related corotation torque is positive and can eventually counteract the effect of the (negative) 
differential Lindblad torque, depending on the local temperature gradient and provided that the viscous and thermal 
diffusion coefficients are large enough to keep the entropy-related corotation torque from saturating. \\
% In the context of planetary synthesis models, {\bf our positive results suggest} that taking into account the entropy-related 
%corotation torque may help to reduce discrepancies between these models and observational data without 
%artificially lowering Type I migration rates.\\
The positive results obtained in this paper bring further support for the entropy-related corotation torque as a possible solution for the discrepancy between the mass-semi-major axis diagram of observed exoplanets, and that inferred from theories of planetary formation and evolution (e.g., Ida \& Lin 2008, Mordasini et al. 2009, Howard et al. 2010, Hellary \& Nelson 2012).\\
The simulations we have presented have neglected radiative processes and we will present in a future paper 
simulations  with radiative transfert treated in the flux-limited approximation. We will also investigate the effects of turbulence on the dynamics of 
multiple protoplanets located in the vicinity of an opacity transition where an entropy jump can act as a planet trap
and promotes embryo growth (Horn et al. 2012).  
%\begin{acknowldgements}
%Simulations were performed using HPC resources from GENCI-cines (c2011026735).
%\end{acknowldgements}
\section*{Acknowledgments}
Simulations were performed using HPC resources from GENCI-cines (c2012046733). Cl\'ement Baruteau is 
supported by a Herchel-Smith Postdoctoral Fellowship of the University of Cambridge.

\label{lastpage}

\end{document}